\documentclass[a4paper,10pt,notitlepage]{article}
\usepackage{amsfonts,amssymb,amsmath,amsthm,hyperref,colonequals}
\usepackage{caption}
\usepackage{cite}
\usepackage{tabularx}
\usepackage{graphicx}
\usepackage{shuffle}
\usepackage{esint}
\usepackage[nice]{nicefrac} 
\usepackage{url}
\usepackage{bm}
\usepackage{lmodern}
\theoremstyle{definition}
\newcommand{\beqa}{\begin{eqnarray}}
\newcommand{\eeqa}{\end{eqnarray}}
\newcommand{\beq}{\begin{equation}}
\newcommand{\eeq}{\end{equation}}

\newcommand{\SU}[1]{\ensuremath{\text{SU}\left({#1}\right)}}

\newcommand{\calH}{\mathcal{H}}
\newcommand{\calS}{\mathcal{S}}

\newcommand{\calO}{\mathcal{O}}
\newcommand{\calZ}{\mathcal{Z}}
\newcommand{\calN}{\mathcal{N}}
\newcommand{\calW}{\mathcal{W}}
\newcommand{\calC}{\mathcal{C}}

\newcommand{\resa}{r_\chi}
\newcommand{\resb}{s_\chi}
\newcommand{\moment}{\mathfrak{m}}
\newcommand{\car}{\hat{\textbf{a}}}

\newcommand{\parf}{Z}
\newcommand{\nv}{r}
\newcommand{\app}{P}
\newcommand{\coef}{\textsf{k}}
\newcommand{\Cusp}{\Gamma_{\text{cusp}}}
\newcommand{\cusp}{K}
\newcommand\tr{\text{Tr}}

\newcommand{\vac}[1]{\ensuremath{\left< \, #1\, \right>}}

\begin{document}

\thispagestyle{empty}
\setcounter{page}{0}
\begin{flushright}\footnotesize
\texttt{DESY 15-196}\\
\texttt{HU-Mathematik-2015-12}\\
\texttt{HU-EP-15/50}\\
\texttt{MITP/15-092}\\
\vspace{0.5cm}
\end{flushright}
\setcounter{footnote}{0}

\begin{center}
{\huge{
\textbf{Exact Bremsstrahlung 
and  Effective Couplings
}
}}
\vspace{15mm}

{\sc 
Vladimir Mitev$^{a,b}$,   Elli Pomoni$^{c,d}$ }\\[5mm]

{\it $^a$Institut f\"ur Physik, WA THEP\\
Johannes Gutenberg-Universit\"at Mainz\\
Staudingerweg 7, 55128 Mainz, Germany
}\\[3mm]

{\it $^b$Institut f\"ur Mathematik und Institut f\"ur Physik,\\ Humboldt-Universit\"at zu Berlin\\
IRIS Haus, Zum Gro{\ss}en Windkanal 6,  12489 Berlin, Germany
}\\[3mm]

{\it $^c$DESY Hamburg, Theory Group, \\
Notkestrasse 85, D--22607 Hamburg, Germany
}\\[3mm]

{\it $^d$Physics Division, National Technical University of Athens,\\
15780 Zografou Campus, Athens, Greece
}\\[3mm]

\texttt{vmitev@uni-mainz.de}\\
\texttt{elli.pomoni@desy.de}\\[10mm]

\textbf{Abstract}\\[2mm]
\end{center}
We calculate supersymmetric Wilson loops on the ellipsoid for a large class of $\calN=2$ SCFT using the localization formula of Hama and Hosomichi.
From them we extract the radiation emitted by an accelerating  heavy probe quark as well as the entanglement entropy following the recent works of Lewkowycz-Maldacena and Fiol-Gerchkovitz-Komargodski. Comparing our results with the $\calN=4$ SYM ones, we obtain  interpolating functions $f(g^2)$
such that a given $\calN=2$ SCFT observable is obtained by replacing in the corresponding
$\calN=4$ SYM result the coupling constant by $f(g^2)$. 
These ``exact effective couplings'' encode the finite, relative renormalization between the $\calN=2$ and the $\calN=4$ gluon propagator and  they interpolate between the weak and the strong coupling. We discuss the range of their applicability.


\newpage
\setcounter{page}{1}


\tableofcontents
\addtolength{\baselineskip}{5pt}

\section{Introduction}

Thanks to its maximal supersymmetry, $\mathcal{N}=4$ SYM is the best understood interacting gauge theory in four dimensions. It behooves us to apply our knowledge of  $\mathcal{N}=4$ SYM to extract results for other gauge theories. We recently observed that some quantities in certain gauge theories can be obtained by a ``substitution rule'' in which the $\mathcal{N}=4$ gauge coupling is replaced by an \textit{effective coupling} \cite{Pomoni:2013poa,Mitev:2014yba}. The essential question then arises: are the effective couplings universal, {\it i.e.} independent of the observable computed? Addressing this question is the main motivation of this paper.

A powerful tool in the study of $\mathcal{N}=4$ SYM is integrability  \cite{Beisert:2010jr}. Integrability was also discovered in the spectral problem of planar $\calN=2$ SCFTs, for a purely gluonic subset of local operators  with $SU(2,1|2)$ symmetry that is closed under renormalization \cite{Pomoni:2013poa}.
 The mixing matrix of anomalous dimensions of planar $\calN=2$ SCFTs is obtained by the $\calN=4$ SYM result after replacing the  $\calN=4$ SYM coupling constant $g^2$ by an effective coupling $f(g^2_i)$, a function of all the marginal couplings $g^2_i$ of the $\calN=2$ SCFT, computed via localization in \cite{Mitev:2014yba}.

Pestun's work on localization \cite{Pestun:2007rz} has led to a plethora of exact results for gauge theories in four dimensions with $\calN\geq 2$ supersymmetry, see \cite{Teschner:2014oja} for a  review. These include the vacuum expectation values of supersymmetric Wilson loops and 't Hooft Loops \cite{Pestun:2007rz,Pestun:2009nn,Gomis:2011pf} as  well as other observables, not immediately given by localization, such as	
the cusp anomalous dimension, the entanglement entropy\footnote{The traditional definition of  entanglement entropy comes from thermal field theory where the antisymmetric boundary conditions for the fermions break supersymmetry completely.}  \cite{Lewkowycz:2013laa} and   
the quark anti-quark potential \cite{Drukker:2011za,Correa:2012hh,Drukker:2012de}.\footnote{The  Zamolodchikov metric given by the two-point functions of the exactly marginal operators is another  very interesting 
 non-BPS observable that can be extracted from exact localization results \cite{Gerchkovitz:2014gta,Gomis:2015yaa}.}

\begin{figure}[t]
 \centering
  \includegraphics[height=3.5cm]{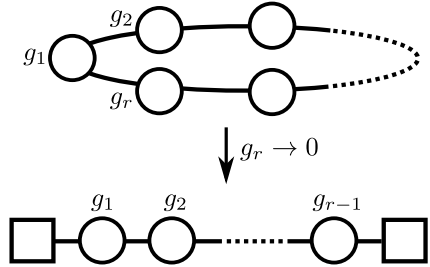}
  \caption{\it The $\hat{A}_{\nv-1}$ $\calN=2$ SCFT  elliptic quivers with $SU(N)^\nv$ color group.  Linear quiver theories can  be obtained by taking a limit in which one of the gauge couplings $g_r\rightarrow 0$. This procedure produces the correct results only in the weak coupling expansion. The strong coupling limit and $g_r\rightarrow 0$ do not commute.}
  \label{fig:QuiverTheories}
\end{figure}

Searching for more observables $O$ to which the coupling substitution rule can be applied, we 
 compute via localization the large $N$ limit of the $b$-deformed  BPS Wilson loops of  \cite{Hama:2012bg} for a large class\footnote{Obtaining similar results for other  $\calN=2$ SCFT with a Lagrangian description is straightforward. For theories that do not have a Lagrangian description, localization is not applicable, and thus the road is not completely paved yet. However, we believe that such Wilson loops could be obtained relatively straightforward, by using the results of \cite{Bao:2013pwa,Hayashi:2013qwa,Mitev:2014isa,Isachenkov:2014eya} as well as  by combining with AGT intuition.}
  of $\calN=2$ SCFT,
 the quiver diagram of which is depicted in figure \ref{fig:QuiverTheories}.
From them we extract following  \cite{Lewkowycz:2013laa,Fiol:2015spa} the entanglement entropy and the radiation emitted by an accelerating  heavy probe quark.
We compare these observables $O$ with their $\mathcal{N}=4$ SYM counterparts, extract the effective couplings $f_{O}$
and address the question of universality.
 Our results  interpolate between the weak and the strong coupling. From the weak coupling we can understand the first few terms in the expansion of the localization result using Feynman diagrams,
  while from the strong coupling the leading term using AdS/CFT.

This paper is structured as follows. We begin in section \ref{sec:review} with a review of the Bremsstrahlung function and the integrability of the purely gluonic sector in $\calN=2$ SCFTs. We then overview in section \ref{sec:Hosomichi} the setup of the circular Wilson loops on the ellipsoid and their computation via localization. We follow up in section \ref{sec:saddlepoint} with the saddle point approximation in the planar limit.  We discuss the results on  Bremsstrahlung function and the entanglement entropy in section \ref{sec:bremsstrahlung}.
The technical  aspects of the weak and strong coupling solutions to the  saddle point equations are kept in the appendices \ref{app:weakcouplingexpansion} and \ref{app:strongcouplinglimit}.
We present an interpretation of some aspects of results, in particular the universality of the coupling substitution in section \ref{universality}. 
Finally, we conclude and make some suggestions for future work in section \ref{sec:conclusions}.

\section{Review}
\label{sec:review}
 
In this section, we present for the convenience of the reader a short review of the main ingredients appearing in this paper. We introduce the cusp anomalous dimension as well as the Bremsstrahlung function. We then explain how to obtain the Bremsstrahlung function in $\calN=2$ SCFTs using localization and the work of \cite{Fiol:2015spa}. Finally, we review our previous work on the coupling substitution rule in the spectral problem of the purely gluonic sector of $\calN=2$ SCFTs and show how it can be computed.
 
 \subsection{The cusp anomalous dimension and the   Bremsstrahlung function}
 
\begin{figure}[h]
 \centering
  \includegraphics[height=1.5cm]{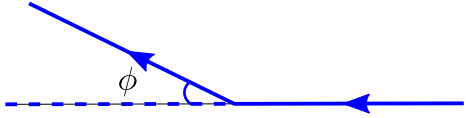}
  \caption{\it The Wilson line with an Euclidean cusp angle $\phi$. }
  \label{fig:CuspedLine2}
\end{figure}

 The energy emitted by an uniformly accelerating probe quark is proportional to the Bremsstrahlung function $B$
\beq
\Delta E=2\pi B \int dt \dot{v}^2\,,
\eeq
for small velocities $v$.
It is well known, see for example \cite{Brock:1993sz} for an old review and \cite{Correa:2012at} for a more recent presentation, that $B$ can be obtained from a Wilson line that makes a sudden turn by an angle $\phi$, see figure~\ref{fig:CuspedLine2}, at a single specific point that we refer to as the cusp. 
As discussed already in \cite{Polyakov:1980ca}, the vacuum expectation value of such a Wilson loop is both UV and IR divergent with divergences of the form
\beq
\label{eq:Wilson}
\vac{W_{\varphi}}\sim e^{-\Cusp(\varphi)\log \frac{\Lambda_{\text{UV}}}{\Lambda_{\text{IR}}}}\,,
\eeq
where $\Lambda_{\text{UV}}$, respectively $\Lambda_{\text{IR}}$ is the UV, respectively IR cutoff.  In \eqref{eq:Wilson}, we have analytically continued to Minkowski signature by setting $\phi=i\varphi$.

The first important observation for this Wilson loop is the fact that, for large Euclidean angles $\phi\sim \pi$,   its  cusp anomalous dimension leads to the quark-antiquark potential, while for large values of the Minkowski angle $\varphi$, it grows linearly with $\varphi$ with a slope equal to the light-like cusp anomalous dimension $\cusp$, {\it i.e.}
\beq
\Cusp(\varphi)\sim \cusp\varphi\,.
\eeq
The number $ \cusp$,  determines the leading logarithmic  behavior of the anomalous dimensions of finite twist operators in the large spin limit  \cite{Gubser:2002tv} as
\beq
\label{cuspTwistOperators}
\Delta -S \sim  \cusp \log(S)\, .
\eeq
The light-like cusp anomalous dimension has been calculated for $\mathcal{N}=4$ SYM to four loops \cite{Correa:2012nk,Henn:2012qz,Henn:2013wfa}, also using integrability \cite{Beisert:2006ez}, see \cite{Freyhult:2010kc} for a review, and lately even using resurgence techniques  \cite{Dorigoni:2015dha, Aniceto:2015rua}.

The second important observation is that, for small $\varphi$, the divergence of the Wilson loop becomes quadratic in $\varphi$ with coefficient given by the  Bremsstrahlung function
\beq
\label{eq:definitionB}
\Cusp(\varphi)=B\varphi^2+\mathcal{O}(\varphi^4)\,.
\eeq
Importantly, the Bremsstrahlung function can be obtained from other geometries that allow us to compute it via localization. In \cite{Correa:2012at}, it was argued that for $\calN=4$ SYM, $B$ can be obtained from the Wilson loop expectation value on the sphere as $B=\frac{\lambda}{2\pi} \partial_{\lambda}\log \langle W\rangle$.\footnote{Here and elsewhere, we use the following definition of the couplings
$\lambda=N g_{\text{YM}}^2=(4\pi g)^2$.
}
Furthermore,  according to a conjecture by \cite{Fiol:2015spa}, the Bremsstrahlung function of $\calN=2$ theories should be given by 
\beq
\label{eq:bremsstrahlungfiol}
B=\pm \frac{1}{4\pi^2}\frac{d}{db}\log \langle W^{\pm}(b)\rangle_{\big|b=1}\,,
\eeq
where $\langle W^{\pm}(b)\rangle$ are the Wilson loop expectation values of circular loops on the ellipsoid with parameter $b$, see \eqref{eq:WilsonloopHosomichi}.  
This formula is true for $\mathcal{N}=4$ SYM \cite{Lewkowycz:2013laa} but needs to be subjected to further checks or to be derived rigorously for $\mathcal{N}=2$ theories. Formula \eqref{eq:bremsstrahlungfiol} is an input of our paper. However, our study provides consistency checks of \eqref{eq:bremsstrahlungfiol} by verifying that it is compatible with the coupling substitution rule, both in the weak and in the strong coupling limit.

\subsection{The substitution rule in the  purely gluonic $SU(2,1|2)$ sector}

In  \cite{Pomoni:2013poa},\footnote{See \cite{Korchemsky:2010kj,Beisert:2004fv,Gadde:2009dj,Gadde:2010zi,Gadde:2010ku,Pomoni:2011jj,Liendo:2011xb,Gadde:2012rv} for work on which it was based.} 
integrability was discovered in the spectral problem of planar $\calN=2$ SCFTs  for a purely gluonic subset of local operators, closed under renormalization. This set of operators is made out of the fields $\phi,\lambda_+^{\mathcal{I}}, \mathcal{D}_{+ \dot{\alpha}}$ in one of the $\mathcal{N}=2$ vector multiplets of the theory and transforms under the action of a $SU(2,1|2)$ global symmetry.  
The mixing matrix of anomalous dimensions\footnote{The planar limit is essential at this stage for two reasons. Firstly, in order to use the integrability of $\calN=4$ SYM, we have to go to the planar limit. Secondly, the inheritance theorems of \cite{Bershadsky:1998mb,Bershadsky:1998cb}, according to which the correlation function in the untwisted sector are equal to the $\calN=4$ ones at the orbifold point,  hold only in the planar limit.} of planar $\calN=2$ SCFTs is obtained by the $\calN=4$ SYM result after replacing the  $\calN=4$ SYM coupling constant as $g^2\rightarrow f(g^2_i)$, with $f$ a function of all the marginal couplings $g^2_i$ of the $\calN=2$ SCFT.
To be more precise, there is a purely gluonic sector together with its respective effective coupling $f_k(g_1^2,\ldots, g_\nv^2)$  for each vector multiplet $V_k$ of the theory, but for the $A$-type quivers that we study, the effective couplings are all related by permutations of the marginal couplings. 
In particular, the light-like cusp anomalous dimension $\cusp$ is an observable in the purely gluonic sector.  As we already mentioned above, $ \cusp$  is the leading logarithmic behavior of the anomalous dimensions of finite twist $\Delta-S$ operators in the large spin limit \eqref{cuspTwistOperators}. Thus, it
is in the purely gluonic $SU(2,1|2)$  sector, since  the twist $\Delta-S$ operators are in the sector. Thus, for $\calN=2$ SCFTs the  light-like cusp anomalous dimension $\cusp_{\calN=2,k}$ for the $k^{\text{th}}$ gauge group is simply given by the $\calN=4$ results by the substitution
\beq
\label{N2cuspfromN4}
\cusp_{\calN=2,k}(g_i^2)\equiv \cusp_{k}(g_i^2)=\cusp_{\calN=4}(f_k(g^2_i)) \, .
\eeq

 From the Feynman diagrams (weak coupling) point of view, the effective couplings $f_k(g^2_i)$ 
 is the relative finite renormalization of the   $\calN=2$ gluon propagator of the $k^{\text{th}}$ color group \cite{Mitev:2014yba},
\beq
\label{RelativeFiniteRenorm}
f_k(g_1^2,\ldots, g_\nv^2)=g_k^2+g_k^2\left[\left(\calZ_{g_k}^{\calN=2}\right)^2-\left(\calZ_{g_k}^{\calN=4}\right)^2\right].
\eeq
and as such depends on all the marginal couplings $g_i^2$ of the $\calN=2$ SCFT.
In  \cite{Mitev:2014yba}, we computed using Feynman diagrams the  relative finite renormalization of the  $\calN=2$ gluon propagator to three-loops and found that
\begin{equation}
\label{4loopPropagator}
f_k(g_1^2,\ldots, g_\nv^2)=g_k^2+6\zeta(3) g_k^4 \left[g_{k-1}^2+g_{k+1}^2-2 g_k^2\right] -20 \zeta(5) g_k^4 \left[g_{k-1}^4+g_{k+1}^4-6 g_k^4+2 g_k^2 \left(g_{k-1}^2+g_{k+1}^2\right)\right]  \, .
\end{equation}

From the strong coupling point of view and using AdS/CFT correspondence\footnote{Note that the quivers that we are considering have a gravity dual description.}, the effective coupling computes the relation (via the AdS/CFT  dictionary) between the effective tension of the string and the coupling constant of the  $\calN=2$ SCFT
\begin{equation}
\label{AdSdictionaty}
T_{\text{eff}}^2= \frac{R^4}{(2 \pi \alpha')^2} = f(g_i^2)  \,.
\end{equation}
Using the AdS/CFT   dictionary and the work of \cite{Lawrence:1998ja,Gadde:2009dj, Gadde:2012rv} in \cite{Mitev:2014yba}, we obtained  the leading term of the effective tension of the string at strong coupling and found that it is
\begin{equation}
\label{AdSresult}
f_k(g_i^2)=r\frac{g_1^2\cdots g_r^2}{\sum_{i=1}^r\prod_{j\neq i}g_j^2}+\cdots,
\end{equation}
 On the string theory side, the observables of the purely gluonic $SU(2,1|2)$ sector correspond to string states classically living in the $AdS_5\times S^1$ factor\footnote{Specifically, the geometry does not factorize, but has an $U(1)$ isometry.} of the geometry with the $S^1$ corresponding to the $U(1)_r$ of the $\calN=2$ theories. The chiral $\mbox{Tr} (\phi^{\ell})$ with $\Delta = r$ are charged under the $U(1)_r$ and correspond to sugra KK reduction modes on this $S^1$. See \cite{Gadde:2009dj} and also \cite{Aharony:2015zea}  for a recent discussion.

In \cite{Mitev:2014yba}  the effective couplings  were also extracted from the circular Wilson loop expectation value on $S^4$, calculated thanks to localization \cite{Pestun:2007rz}. The results agree with \eqref{4loopPropagator} and \eqref{AdSresult}. It now is vital to study as many observables as possible in order to check the extend to which the effective couplings are universal. 

 A short comment about notation is due. Since we will often be comparing $\calN=4$ quantities to $\calN=2$ ones, we need to be painstakingly clear about denoting them properly. In general $\calN=4$ quantities will be denoted as such, for example $\langle W_{\calN=4}\rangle$ for the vacuum expectation value of the Wilson loop on the sphere.
 We will specifically be considering the $\mathbb{Z}_\nv$ cyclic quiver $\calN=2$ theories, see figure \ref{fig:QuiverTheories}. They have $\nv$ gauge groups and we will designate the corresponding quantities, such as the Wilson loop expectation values or the Bremsstrahlung functions, simply by labeling them by an index $k\in \{1,\ldots, \nv\}$, for example $B_k$. Furthermore, for the sake of brevity, we shall often abbreviate the dependence of a function $f_k(g_1^2,\ldots, g_\nv^2)$ of all the couplings as $f_k(g_i^2)$.

\section{Wilson loops on Ellipsoids}
\label{sec:Hosomichi}

The 4D ellipsoid is defined, in embedding coordinates, via the equation 
\beq
\frac{x_0^2}{r^2}+\frac{x_1^2+x_2^2}{\ell^2}+\frac{x_3^2+x_4^2}{\tilde{\ell}^2}=1\,.
\eeq
\begin{figure}[h]
 \centering
  \includegraphics[height=3.5cm]{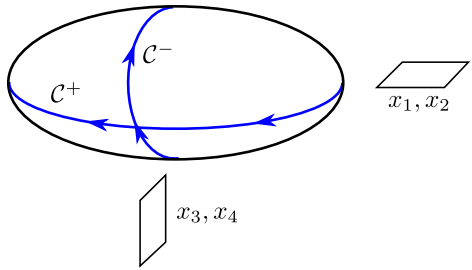}
  \caption{\it We sketch here two circular Wilson loops $\mathcal{C}^{\pm}$ that can be computed on the ellipsoid via localization. The 4D ellipsoid itself should be understood as a fibration of the 3D ellipsoid over the interval $x_0\in [-r,r]$.}
  \label{fig:WilsonEllipsoid}
\end{figure}
We are interested in  the Maldacena-Wilson loops on the ellipsoid
\begin{equation}
\label{eq:wilsonloopexpectationvalue}
\langle W_{k}(\mathcal{C}^{\pm})\rangle\equiv \langle W_{k}^{\pm}\rangle=\vac{\frac{1}{N}\tr_{\square}\text{Pexp}\oint_{\calC^{\pm}}ds \left(i A_{\mu}^{(k)}(x)\dot{x}^{\mu}+\phi^{(k)}(x)|\dot{x}|\right)}\ ,
\end{equation}
where $\square$ denotes the fundamental representation and $\calC^{\pm}$ are the two circular loop located depicted in figure~\ref{fig:WilsonEllipsoid}. In our case, we have $\nv$ gauge groups and the index $k$ labels the adjoint scalar $\phi^{(k)}$  and the gauge field $A_{\mu}^{(k)}$ in the vector multiplet of the $k$-th gauge group.

We define the deformation parameter $b$ as
\beq
b\colonequals \sqrt{\nicefrac{\ell}{\tilde{\ell}}}\,.
\eeq
The case $b=1$ corresponds to the sphere $S^4$, in which case the localization result was already given by \cite{Pestun:2007rz}. 
In \cite{Hama:2012bg}, the following expression for the partition function (written for simplicity here for a single $\SU{N}$ gauge group) was given
\beq
\label{eq:partitionfunctionHosomichi}
\parf=\int d\hat{a}e^{-\frac{8\pi^2}{g_{\text{YM}}^2}\tr(\hat{a}^2)}\parf_{\text{1-loop}}(a,b)\left|\parf_{\text{inst}}(a,b)\right|^2\,,
\eeq
where $\hat{a}\colonequals \sqrt{\ell \tilde{\ell}}a$. The matrix $a=\text{diag}(a_1,\ldots,a_N)$ is subject to the condition $\sum_{i=1}^Na_i=0$ and is an element of the Cartan subalgebra of $\mathfrak{su}(N)$. The two Wilson loops that we are able to compute are drawn in figure~\ref{fig:WilsonEllipsoid} and are given by
\beq
\label{eq:WilsonloopHosomichi}
\langle W^{\pm}(b)\rangle= \frac{1}{\parf}\int d\hat{a}\tr\left(e^{-2\pi b^{\pm 1}\hat{a}}\right)e^{-\frac{8\pi^2}{g_{\text{YM}}^2}\tr(\hat{a}^2)}\parf_{\text{1-loop}}\left|\parf_{\text{inst}}\right|^2
\eeq
The perturbative part  $Z_{\text{1-loop}}$ of \eqref{eq:partitionfunctionHosomichi} is given by a product over the vector multiplet and hypermultiplet contributions (here only in the bifundamental representation of two $\SU{N}$ gauge groups) with
\beq
\label{eq:1loopHosomichi}
\begin{split}
Z^{\text{vect}}_{\text{1-loop}}&=\prod_{i<j=1}^N\Upsilon\big(i\sqrt{\ell \tilde{\ell}}(a_i-a_j);b\big)\Upsilon\big(-i\sqrt{\ell \tilde{\ell}}(a_i-a_j);b\big)\\
Z^{\text{hyper}}_{\text{1-loop}}&=\prod_{i,j=1}^N\Upsilon\big(i\sqrt{\ell \tilde{\ell}}(a_i^{(1)}-a_j^{(2)})+\nicefrac{Q}{2};b\big)^{-1}\,,
\end{split}
\eeq
where  $Q=b+b^{-1}$ and $\Upsilon$ is defined in appendix C of  \cite{Mitev:2014isa}. We remark that both $Z^{\text{vect}}_{\text{1-loop}}$ and $Z^{\text{hyper}}_{\text{1-loop}}$ are invariant under the transformation $b\leftrightarrow b^{-1}$.

For massless theories, we can rescale the integration variable $a$ and get rid of the factor of $\sqrt{\ell \tilde{\ell}}$, whereas for massive one, the factor of $\sqrt{\ell \tilde{\ell}}$ is part of the ambiguity of the mass. Rearranging some factors in the special functions, leads us to the expression, proven in appendix \ref{app:rewritingthepartitionfunction},
\beq
\label{eq:partitionfunctionHosomichi2}
\parf=\int dae^{-\frac{N}{2g^2}\sum_{i=1}^Na_i^2}\parf_{\text{1-loop}}(a,b)\left|\parf_{\text{inst}}(a,b)\right|^2\,,
\eeq
with $g$ given through the relations $\lambda=N g_{\text{YM}}^2=(4\pi g)^2$ and where now the 1-loop part is given by 
\beq
\label{eq:1loopHosomichiVersion2}
\begin{split}
\parf^{\text{vect}}_{\text{1-loop}}&=\prod_{i<j=1}^N(a_i-a_j)^2\prod_{i,j=1}^NH_v(a_i-a_j;b)\,,\qquad 
\parf^{\text{hyper}}_{\text{1-loop}}=\prod_{i,j=1}^NH_h(a_i^{(1)}-a_j^{(2)};b)^{-1}\,,
\end{split}
\eeq
with the functions $H_v(x;b)$ and $H_h(x;b)$ defined in \eqref{eq:definitionHvandHh}. These functions have the advantage of being simpler to work with for the weak coupling expansion since they are even in $x$, invariant under $b\leftrightarrow b^{-1}$ and normalized as $H_v(0;b)=H_h(0;b)=1$.

\section{Saddle point approximation }
\label{sec:saddlepoint}

We only consider the $A$ class of $\calN=2$ SCFTs, {\it i.e.} the $\mathbb{Z}_r$ cyclic quivers or the linear quivers.\footnote{Similar calculations can be found in \cite{Passerini:2011fe,Russo:2012ay,Russo:2013kea,Russo:2013sba}, and for the $\calN=2^*$ theory on the ellipsoid  in \cite{Marmiroli:2014ssa}.} We shall concentrate on the cyclic, or class $\hat{A}_{\nv-1}$, of quiver theories, since in the weak coupling limit we can recover the results for the class $A_{\nv-1}$ linear quivers by taking a limit, see figure~\ref{fig:QuiverTheories}.  
The partition function is can then be written as 
\beq
\begin{split}
\parf&=\int\prod_{k=1}^\nv d^{N-1}a^{(k)}\prod_{i<j=1}^{N}\left(a_i^{(k)}-a_j^{(k)}\right)^2e^{-\frac{N}{2g_{k}^2}\sum_{i=1}^{N}\left(a_i^{(k)}\right)^2}\parf_{\text{1-loop}}\left|\parf_{\text{inst}}\right|^2=\int \prod_{k=1}^\nv d^{N-1}a^{(k)}e^{-N \calS_{\text{eff}}}\,,
\end{split}
\eeq
where the instanton part $\parf_{\text{inst}}$ will be ignored, since we are going to perform a planar limit\footnote{See \cite{Passerini:2011fe} and \cite{Russo:2012ay} for a discussion on this subject.}  computation. We remind that, since we are dealing with $SU(N)$ gauge groups, the Coulomb parameters satisfy the equation
$\sum_{i=1}^{N}a_i^{(k)}=0\,, \forall k=1,\ldots, \nv $.
In the large $N$ limit that we are interested in, we can safely ignore the instanton part \cite{Passerini:2011fe}, and the effective action is given by
\beq
\calS_{\text{eff}}=\sum_{k=1}^\nv\left[\sum_{i=1}^{N}\frac{1}{2g^2_k}\left(a_i^{(k)}\right)^2-\frac{1}{N}\sum_{i<j=1}^{N}\ln\left(a_i^{(k)}-a_j^{(k)}\right)^2\right]-\frac{1}{N}\ln\left(\parf_{\text{1-loop}}\right).
\eeq
For the $\hat{A}_{\nv-1}$ quivers,  the perturbative part of the partition function can be written as
\beq
\ln\left(\parf_{\text{1-loop}}\right)=\sum_{k,l=1}^\nv\sum_{i,j=1}^N\left[\delta_{kl}\log H_v(a_i^{(k)}-a_j^{(l)})-\frac{\delta_{k,l+1}+\delta_{k,l-1}}{2}\log H_h(a_i^{(k)}-a_j^{(l)})\right]\,.
\eeq
Solving the matrix model in the planar limit is done by considering the saddle point approximation. Specifically, we are computing a vacuum expectation value of some quantity $W$ in the large $N$ limit
\beq
\label{eq:vacexpplanarlimit1}
\vac{W}=\frac{\int \prod_{k=1}^rd^{N-1}a^{(k)}W(a^{(\ell)})e^{-N\calS_{\text{eff}}}}{\int \prod_{k=1}^rd^{N-1}a^{(k)}e^{-N\calS_{\text{eff}}}}\,,
\eeq
where by abuse of notation $W$ is also a function of the eigenvalues $a^{(k)}=(a_1^{(k)},\ldots, a_N^{(k)})$ of the $k^{\text{th}}$ gauge group.
Let the effective action have an extremum ($\frac{\partial \calS_{\text{eff}}}{\partial a}_{\big|a=b}=0$) at $a=(a^{(1)},\ldots, a^{(\nv)})\equiv b$ and expand the integral $a=b+\frac{1}{\sqrt{N}}x$, we get
\beq
\begin{split}
\int daW(a)e^{-N\calS_{\text{eff}}}=&\frac{e^{-N\calS_{\text{eff}}(b)}}{N^{\text{power}}}\int dx e^{-\frac{1}{2}(\calS_{\text{eff}})_{,ij}(b)x_ix_j}\Bigg[W(b)+\\&+\frac{1}{N^{\frac{1}{2}}}\left(W_{,i}(b)x_i-\frac{1}{6}W(b)(\calS_{\text{eff}})_{,ijk}x_ix_jx_k\right)+\cdots\Bigg]\,,
\end{split}
\eeq
where ``power'' is a number that will drop out in the end and we have used the shorthand $W_{,ijk\cdots}=\partial_i\partial_j\partial_k\cdots W$. It follows that the leading term in the planar limit is given by the function evaluated at the saddle point, {\it i.e.}
\beq
\label{eq:vacexpplanarlimit2}
\begin{split}
\vac{W(a)}&=W(b)+\mathcal{O}(\nicefrac{1}{N})\,.
\end{split}
\eeq
In our case, the saddle point equations 
$\nicefrac{\partial \calS_{\text{eff}}}{\partial a_i^{(k)}}=0$
 imply that for all $i=1,\ldots, N$ and all $k=1,\ldots, \nv$, we must have
\beq
\label{eq:saddlepoint1}
\begin{split}
\frac{a_i^{(k)}}{2g_k^2}=&\frac{1}{N}\sum_{j\neq i}\frac{1}{a_i^{(k)}-a_j^{(k)}}-\frac{1}{N}\sum_{l=1}^\nv\sum_{j=1}^N\left[\delta_{kl}K_v(a_i^{(k)}-a_j^{(l)})-\frac{\delta_{k,l+1}+\delta_{k,l-1}}{2} K_h(a_i^{(k)}-a_j^{(l)})\right]\,.
\end{split}
\eeq
where $K_v$ and $K_h$ are defined in \eqref{eq:definitionsKvandKh}.
In the $N\rightarrow \infty$ limit, we replace the eigenvalues $a_i^{(k)}$ by normalized densities that are localized on a symmetric interval $[-\mu_k,\mu_k]$ 
\beq
\label{eq:rhodefandnormalization}
\rho_k(x)=\frac{1}{N}\sum_{i=1}^N\delta\left(x-a_i^{(k)}\right)\,,\qquad\int_{-\mu_k}^{\mu_k}\rho_k(x)dx=1\,, 
\eeq
which transforms the saddle point equations \eqref{eq:saddlepoint1} into integral equations. Specifically, we obtain the following system of coupled integral equations:
\beq
\label{eq:saddlepoint2}
\frac{x}{2g^2_k}=\fint_{-\mu_k}^{\mu_k}dy\frac{\rho_k(y)}{x-y}-\sum_{l=1}^\nv\int_{-\mu_l}^{\mu_l}\left[\delta_{kl}K_v(x-y)-\frac{\delta_{k,l+1}+\delta_{k,l-1}}{2} K_h(x-y)\right]\rho_l(y)dy
\eeq
for $k=1,\ldots, \nv$. 
For numerical approximations at small values of the 't Hooft couplings, it is sometimes helpful to rewrite \eqref{eq:saddlepoint2} by inverting the Hilbert kernel, {\it i.e.} by acting with $\fint_{-\mu_k}^{\mu_k}\frac{dx}{\sqrt{\mu_k^2-x^2}}\frac{1}{z-x}$ on both sides of the equation. Using \eqref{eq:simpleintegralidentities} and \eqref{eq:invertHilbertkernel} we get the set of equations
\beq
\label{eq:saddlepoint3}
\begin{split}
\rho_k(x)=&\frac{1}{2\pi g_k^2}\sqrt{\mu_k^2-x^2}-\frac{1}{\pi^2}\fint_{-\mu_k}^{\mu_k}\frac{dy}{x-y}\sqrt{\frac{\mu_k^2-x^2}{\mu_k^2-y^2}}\sum_{l=1}^\nv\int_{-\mu_l}^{\mu_l}\rho_l(z)dz\Big[\delta_{kl}K_v(y-z)\\&-\frac{\delta_{k,l+1}+\delta_{k,l-1}}{2} K_h(y-z)\Big]\,,
\end{split}
\eeq
subject to the normalization condition for the densities
\beq
1=\frac{\mu_k^2}{4g_k^2}+\frac{1}{\pi}\int_{-\mu_k}^{\mu_k}\frac{dy y}{\sqrt{\mu_k^2-y^2}}\sum_{l=1}^\nv\int_{-\mu_l}^{\mu_l}\rho_l(z)dz\Big[\delta_{kl}K_v(y-z)-\frac{\delta_{k,l+1}+\delta_{k,l-1}}{2} K_h(y-z)\Big]\,.
\eeq
Thanks to \eqref{eq:vacexpplanarlimit2}, having obtained the densities by solving the saddle point equations \eqref{eq:saddlepoint3}, we can compute the Wilson loop expectation values for the $k$-th gauge group by plugging the densities 
\beq
\label{eq:Wilsonpmdefinition}
W_k^{\pm}(g_1,\ldots, g_\nv;b)=\vac{\frac{1}{N}\sum_{i=1}^Ne^{-2\pi a_i^{(k)}b^{\pm1}}}=\int_{-\mu_k}^{\mu_k}\rho_k(x)e^{-2\pi xb^{\pm1}}dx\,.
\eeq

\subsection{Weak coupling results}
\label{subsec:weakcouplingresults}

By \textit{weak coupling}, we understand the regime for which \textit{all} the couplings are small, {\it i.e.} 
\beq
\label{eq:scalingofthecouplings}
g_i^2=t\kappa_i\,,
\eeq
with the coefficients $\kappa_i$ being order one constants and $t<<1$.
Appendix \ref{app:weakcouplingexpansion} contains further details on the weak coupling expansion of the Wilson loops. 
From the vacuum expectation values of the Wilson loops on the ellipsoids, we define the ``full'' effective couplings $f_k(g_i^2;b)$ via
\beq
\label{eq:definitionofthefulleffectivecouplings}
\langle W^+_{\calN=4}(f_k(g_1^2,\ldots, g_\nv^2;b);b) \rangle=\langle W^+_{k}(g_1^2,\ldots, g_\nv^2;b) \rangle\,.
\eeq
We could have just as easily used the other Wilson loop $\langle W^-\rangle$ in the above. The corresponding effective coupling is simply $f_k(g_i^2,b^{-1})$.
It is useful to expand these effective couplings in power of $(b-1)$ around $b=1$. 
We define the coefficients of this expansion as 
\beq
f_k(g_1^2,\ldots, g_\nv^2;b)\colonequals \sum_{n=0}^{\infty}f_k^{(n)}(g_1^2,\ldots, g_\nv^2)(b-1)^n\,.
\eeq
In appendix \ref{app:weakcouplingexpansion}, we write explicitly the linear equations that need to be solved to obtain the Wilson loop expectation values on the ellipsoids.
From the result \eqref{eq:resultforWplusbforZ2} for the $\mathbb{Z}_2$ quiver, we get
\beq
\label{eq:effcouplingbforZ2part0}
\begin{split}
f_1^{(0)}(g_1^2,g_2^2)&= g_1^2+\left(g_2^2-g_1^2\right)\Big\{12 \zeta (3)  g_1^4 -40 \zeta (5)  g_1^4 \left[3 g_1^2+g_2^2\right]\\&-\frac{4}{3}  g_1^4  \big[10 \pi ^2 \zeta (5) g_1^4+108 \zeta (3)^2 \left(2 g_1^4-g_2^2 g_1^2+g_2^4\right)-105 \zeta (7) \left(8 g_1^4+5 g_2^2 g_1^2+g_2^4\right)\big]+\\
&-\frac{8}{9} g_1^4 \big[8 \pi ^4 g_1^6 \zeta (5)+21 \pi ^2 \left(11 g_1^2+5 g_2^2\right) g_1^4 \zeta (7)+27 \big(5 g_1^6 (60 \zeta (3) \zeta (5)-91 \zeta (9))\\&-g_2^2 g_1^4 (100 \zeta (3) \zeta (5)+371 \zeta (9))+g_2^4 g_1^2 (20 \zeta (3) \zeta (5)-161 \zeta (9))+g_2^6 (100 \zeta (3) \zeta (5)-21 \zeta (9))\big)\big]\Big\}\\&+\calO(g^{14})\,,
\end{split}
\eeq
for the term constant in $b-1$ and
\beq
\begin{split}
\label{eq:firstderivativeoffgb}
f_1^{(1)}(g_1^2,g_2^2)=&-\left(g_2^2-g_1^2\right)\pi^2\left\{\frac{80}{3} g_1^8   \zeta (5)+\frac{16}{9}  g_1^8 \left(16 \pi ^2 g_1^2 \zeta (5)+21 \left(11 g_1^2+5 g_2^2\right) \zeta (7)\right)\right\}+\calO(g^{14})\,,
\end{split}
\eeq
for the linear piece. In order to not overload the reader with information, we refrain from presenting any additional orders in the $(b-1)$ expansion, since they can be easily taken from \eqref{eq:resultforWplusbforZ2}.

A short remark is in order. The terms in the expansions \eqref{eq:effcouplingbforZ2part0} and \eqref{eq:firstderivativeoffgb}  are homogeneous polynomials in the two couplings $g_1$ and $g_2$ of a given degree. By $\calO(g^{n})$, we mean that the results exclude polynomials of homogeneous degree greater or equal to $n$. 
Lastly, the expression for the other effective coupling $f_2(g_1^2,g_2^2;b)$ can be obtained by using the $\mathbb{Z}_2$ cyclic symmetry of the theory
\beq
f_2(g_1^2,g_2^2;b)=f_1(g_2^2,g_1^2;b)\,.
\eeq
For the general $\mathbb{Z}_\nv$ cyclic quivers, we have the results
\beq
\label{eq:fkzeroforcyclicquivers}
\begin{split}
f_k^{(0)}(g_i^2)&=g_k^2+6\zeta(3) g_k^4 \left[g_{k-1}^2+g_{k+1}^2-2 g_k^2\right]-20 \zeta(5) g_k^4 \left[g_{k-1}^4+g_{k+1}^4-6 g_k^4+2 g_k^2 \left(g_{k-1}^2+g_{k+1}^2\right)\right] \\&+ g_k^4 \Big[70\zeta(7) \Big(g_{k-1}^6+g_{k+1}^6-16 g_k^6+3 g_k^4 \left(g_{k-1}^2+g_{k+1}^2\right)+4 g_k^2 \left(g_{k-1}^4+g_{k+1}^4\right)\Big)\\&-2 \zeta(2)(20\zeta(5)) g_k^4 \left(g_{k-1}^2+g_{k+1}^2-2 g_k^2\right)+(6\zeta(3))^2 \Big(8 g_k^6-2 g_{k-1}^6-2 g_{k+1}^6+g_{k-1}^4 g_{k-2}^2+g_{k+2}^2 g_{k+1}^4\\&-6 g_k^4 \left(g_{k-1}^2+g_{k+1}^2\right)+2 g_k^2 \left(g_{k-1}^4+g_{k-1}^2 g_{k+1}^2+g_{k+1}^4\right)\Big)  \Big]+\mathcal{O}(g^{12})
\end{split}
\eeq
already present in \cite{Mitev:2014yba} and
\beq
\label{eq:firstderivativeoffullfgeneral}
f_k^{(1)}(g_i^2)=80 g_k^8 (2 g_k^2-g_{k+1}^2-g_{k-1}^2) \zeta(2) \zeta (5)+\mathcal{O}(g^{12})\,,
\eeq
for the first $b-1$ correction. The effective couplings for the general cyclic quivers are symmetric under $\mathbb{Z}_\nv$.

\subsection{Strong coupling results}
\label{subsec:strongcouplingresults}

Similarly to the weak coupling approximation in subsection \ref{subsec:weakcouplingresults}, we define the \textit{strong coupling regime} to be the one in which all the couplings are large, {\it i.e.} we suppose that they all scale like \eqref{eq:scalingofthecouplings} with $t>>1$.

In appendix \ref{app:strongcouplinglimit},  we present the strong coupling analysis that we omitted in \cite{Mitev:2014yba}. We obtain namely that, at the leading order, the densities $\rho_k$ behave like
$\rho_k(x)\sim \frac{1}{2\pi g_k^2}\sqrt{\mu_k^2-x^2}$
with the widths
\beq
\label{eq:muinthestrongcoupling}
  \mu_k=\bar{\mu}\colonequals 2\sqrt{\frac{rg_1^2\cdots g_r^2}{\sum_{i=1}^r\prod_{k\neq i}g_k^2}}\qquad \forall k\,.
  \eeq
  This implies that the Wilson loops expectation values asymptotically go like
\beq
\label{eq:strongcouplinglimitWk}
\langle W_k^{\pm}\rangle\sim \frac{ e^{2 \pi  b^{\pm 1} \bar{\mu} }}{2 \pi ^2 b^{\pm 3/2} \bar{\mu}^{\frac{3}{2}}}+\mathcal{O}(b-1)^2\,.
\eeq
The leading piece in the couplings of the above can be written as 
\beq
\label{eq:strongcouplingW}
\log \langle W_k^{\pm} \rangle\sim2 \pi  \mu_k\pm(b-1) \left(2 \pi  \mu_k-\frac{3}{2}\right)+O\left((b-1)^2\right)\,,
\eeq  
up to logarithmic corrections that are sub-leading  and can be dropped.
Due to the exponential term, the strong coupling limit of the effective couplings is simply given by comparing \eqref{eq:strongcouplinglimitofmu} with the width $\mu$ for $\calN=4$. Since for $\calN=4$ SYM we have $\mu^2=4g^2$, comparing with \eqref{eq:strongcouplinglimitofmu}  leads to
\beq
\label{eq:fkinthestrongcoupling}
f_k^{(0)}(g_1,\ldots, g_\nv)=r\Bigg(\sum_{j=1}^r\frac{1}{g_j^2}\Bigg)^{-1}=\frac{rg_1^2\cdots g_r^2}{\sum_{i=1}^r\prod_{k\neq i}g_k^2}\,,
\eeq  
up to constant and logarithmic corrections. Furthermore, to that order of precision $f_k^{(1)}$ is zero, which  can be seen by plugging \eqref{eq:strongcouplingW} into \eqref{eq:definitionofthefulleffectivecouplings},  solving for $f_k$.

\section{The Bremsstrahlung function and the entanglement entropy}
\label{sec:bremsstrahlung}

Having in section \ref{sec:saddlepoint} derived the vacuum expectation value of the Wilson loops on the ellipsoid, it is now time to reap the fruits of our labor and investigate the quantities that we can easily obtain from them, namely the Bremsstrahlung function and the entanglement entropy.

\subsection{The Bremsstrahlung function}

For $\calN=4$, we can obtain the Wilson loop on the ellipsoid by simply making the substitution $g\rightarrow g b^{\pm1}$, leading to the planar limit result
\beq
\label{eq:WilsonN4}
\langle W^{\pm}_{\calN=4}(g^2;b)\rangle=\frac{I_1(4\pi gb^{\pm 1})}{2\pi g b^{\pm 1}}+\calO((b-1)^2)\,,
\eeq
where $I_n$ are the modified Bessel functions of the first kind. 
It follows from \eqref{eq:bremsstrahlungfiol} and \eqref{eq:WilsonN4} that we have in the planar limit the expression (see \cite{Correa:2012at} for an earlier derivation of $B_{\calN=4}$)
\beq
\label{eq:BremsstrahlungN4}
\begin{split}
B_{\calN=4}(g^2)=&\frac{g (I_0(4 \pi  g)+I_2(4 \pi  g))}{2 \pi  I_1(4 \pi  g)}-\frac{1}{4 \pi ^2}=\frac{g I_2(4 g \pi )}{\pi  I_1(4 g \pi )}\,,
\end{split}
\eeq
where we have used $\frac{d}{dx}I_1=\frac{1}{2}(I_0+I_2)$. One can check that for large $g$ we have
$B_{\calN=4}(g)\sim\frac{g}{\pi }-\frac{3}{8 \pi ^2}$.
In particular, $B_{\calN=4}(g)$ is monotonically growing for all $g>0$ and is hence invertible in that domain. It follows that the equation $B_{\calN=4}(x)=y$ has an unique solution for $y$ positive. We now define effective coupling $f_{B;k}$ for the $\mathbb{Z}_\nv$ quiver theories by demanding
\beq
\label{eq:definitionfB}
B_{\calN=4}\big(f_{B;k}(g_1^2,\ldots , g_\nv^2)\big)=B_{k}(g_1^2,\ldots, g_\nv^2)\,.
\eeq
We find that, in the weak coupling $B_{k}$ goes like,
\beq
\begin{split}
B_{k}=&g_{k}^2-\frac{2\pi ^2}{3}  g_{k}^4+\frac{2}{3} g_{k}^4 \Big[\pi ^4 g_{k}^2+9 \zeta (3) \left(-2 g_{k}^2+g_{k+1}^2+g_{k-1}^2\right)\big]\\
&+ \Big[8 \pi ^2 \zeta (3) \left(2 g_{k}^2-g_{k+1}^2-g_{k-1}^2\right) g_{k}^6+20 \zeta (5) \left(6 g_{k}^4-2 \left(g_{k+1}^2+g_{k-1}^2\right) g_{k}^2-g_{k+1}^4-g_{k-1}^4\right) g_{k}^4-\frac{32}{45} \pi ^6 g_{k}^8\Big]\\
&+ \Big[12 \pi ^4 \left(-2 g_{k}^2+g_{k+1}^2+g_{k-1}^2\right) g_{k}^8 \zeta (3)-\frac{40 \pi ^2}{3} \left(10 g_{k}^4-3 \left(g_{k+1}^2+g_{k-1}^2\right) g_{k}^2-2 \left(g_{k+1}^4+g_{k-1}^4\right)\right) g_{k}^6 \zeta (5)\\&+2 g_{k}^4 \Big(16 g_{k}^6 \left(9 \zeta (3)^2-35 \zeta (7)\right)-3 \left(g_{k+1}^2+g_{k-1}^2\right) g_{k}^4 \left(36 \zeta (3)^2-35 \zeta (7)\right)\\&+4 g_{k}^2 \left(g_{k+1}^4 \left(9 \zeta (3)^2+35 \zeta (7)\right)+9 g_{k-1}^2 g_{k+1}^2 \zeta (3)^2+g_{k-1}^4 \left(9 \zeta (3)^2+35 \zeta (7)\right)\right)\\&-36 g_{k-1}^6 \zeta (3)^2+18 g_{k-2}^2 g_{k-1}^4 \zeta (3)^2+18 g_{k+1}^4 g_{k+2}^2 \zeta (3)^2+35 g_{k-1}^6 \zeta (7)+g_{k+1}^6 \left(35 \zeta (7)-36 \zeta (3)^2\right)\Big)\\&+\frac{104}{135} \pi ^8 g_{k}^{10}\Big]+\mathcal{O}\left(g^{12}\right)\,.
\end{split}
\eeq
For the $\mathbb{Z}_2$ quiver, the above can be checked from the explicit result \eqref{eq:resultforWplusbforZ2} for the Wilson loop.

We now wish to discuss the relationships between all the different effective couplings. For the sake of clarity, we shall suppress the indices referring to the gauge groups. From the Wilson loops on the ellipsoids, we extract the effective coupling $f_k(g_i^2;b)$ via \eqref{eq:definitionofthefulleffectivecouplings}.
This defines the $b$-dependent effective coupling $f_k(g_i^2;b)$. For $b=1$, it reduces to the ``Wilson loop'' effective couplings
\beq
f_{W;k}(g_1^2,\ldots, g_\nv^2)\colonequals f_k(g_1^2,\ldots, g_\nv^2;b)_{\big|b=1}\,.
\eeq
that we used in \cite{Mitev:2014yba}. On the other hand, from the Bremsstrahlung function, we can extract $f_{B;k}(g_i^2)$ through \eqref{eq:definitionfB}. Let us see how the two are related. We have
\beq
\begin{split}
B_{k}(g_i^2)&=\frac{1}{4\pi^2}\frac{d}{db}\log \langle W_{k}^+(g_i^2;b)\rangle_{\big|b=1}=\frac{1}{4\pi^2}\frac{d}{db}\log \langle W_{\calN=4}(f_k(g_i^2;b);b)\rangle_{\big|b=1}\\
&=\frac{1}{4\pi^2}\frac{\partial}{\partial g^2}\log \langle W_{\calN=4}(f_{W;k}(g_i^2))\rangle \frac{\partial f_k(g_i^2;b)}{\partial b}_{\big|b=1}+B_{\calN=4}(f_{W;k}(g_i^2))
\end{split}
\eeq
From \cite{Correa:2012at}, we take
\beq
B_{\calN=4}=\frac{\lambda}{2\pi^2} \partial_{\lambda }\log  \langle W_{\calN=4}(\lambda)\rangle \Longrightarrow 
\frac{\partial}{\partial g^2}\log  \langle W_{\calN=4}(g^2)\rangle=\frac{2\pi^2}{g^2}B_{\calN=4}(g^2)\,.
\eeq
Since $B_{\calN=4}(f_{B;k}(g_i^2))\stackrel{!}=B_{k}(g_i^2)$, it follows that 
\beq
\label{BWdiscrepancy}
B_{\calN=4}(f_{B;k}(g_i^2))=\left(1+\frac{f^{(1)}_k(g_i^2)}{2f_{W;k}(g_i^2)}\right)B_{\calN=4}(f_{W;k}(g_i^2))\,,
\eeq
where the first derivative $\partial_b f_k(g_i^2;b)_{\big|b=1}\equiv f^{(1)}_k(g_i^2)$ is given in \eqref{eq:firstderivativeoffgb} for the $\mathbb{Z}_2$ quiver and \eqref{eq:firstderivativeoffullfgeneral} in general.
Hence, the discrepancy between $f_{B;k}$ and $f_{W;k}$ comes from the first derivative of the ``full'' effective coupling $f_k(g_i^2;b)$ at $b=1$. To the order that we care to check, the discrepancies  are always proportional to $\zeta(2n)$.

In the strong coupling limit, the result \eqref{eq:strongcouplinglimitWk} for the Wilson loop expectation values implies that the Bremsstrahlung function for the $k^{\text{th}}$ gauge group goes like
\beq
\label{eq:strongcouplinglimitBk}
B_k\sim\frac{\mu_k }{2 \pi }=\frac{1}{\pi}\sqrt{\frac{rg_1^2\cdots g_r^2}{\sum_{i=1}^r\prod_{k\neq i}g_k^2}}\,,
\eeq
ignoring constant and logarithmic contributions, where the widths $\mu_k$ of the densities are to be found in \eqref{eq:muinthestrongcoupling}. Since the leading contribution to $f^{(1)}_k$ is zero at strong coupling, $f_{B;k}=f_{W;k}$ to the precision we have in that regime.

\subsection{Entanglement entropy}

Combining the results of \cite{Lewkowycz:2013laa,Fiol:2015spa}, for 4D $\calN=2$ SCFTs, the additional entanglement entropy of a spherical region due to the presence of a heavy probe located at its origin is given by
\beq
\label{eq:SasfunctionoflogWandB}
S=\log \langle W\rangle-8\pi^2h_W=\left(1-\frac{2}{3}\partial_b\right)\langle\log W^+\rangle_{\big|b=1}=\log \langle W\rangle-\frac{8\pi^2}{3}B\,.
\eeq
For $\calN=4$, this result combined with expressions \eqref{eq:WilsonN4} and \eqref{eq:BremsstrahlungN4} gives 
\beq
S_{\calN=4}(g^2)=\log \left(\frac{I_1(4 g \pi )}{2 \pi  g}\right)-\frac{8 \pi g}{3}\frac{ I_2(4 g \pi )}{I_1(4 g \pi )}\,.
\eeq
Unlike the Bremsstrahlung function $B_{\calN=4}$ or the Wilson loop expectation $\log \langle W_{\calN=4}\rangle$, the entanglement entropy $S_{\calN=4}$ is not monotonically growing in $g$. Hence, we cannot for general values of the couplings find a single solution to the equation $S_{\calN=4}(f_k(g_i^2))=S_{k}(g_i^2)$, though of course we can do it in the weak or strong coupling limits. We restrict to simply stating the weak and strong coupling expansions of the entanglement entropies by using the results of the appendices \ref{app:weakcouplingexpansion} and \ref{app:strongcouplinglimit}. We find
\beq
\begin{split}
S_k=&-\frac{2\pi ^2 }{3} g_k^2+\frac{10}{9} \pi ^4 g_k^4-\frac{4}{3} \Big[\pi ^2 g_k^4 \left(3 \zeta (3) \left(-2 g_k^2+g_{k+1}^2+g_{k-1}^2\right)+\pi ^4 g_k^2\right)\Big]\\&
+\frac{8\pi ^2}{135}  g_k^4 \Big[-225 \pi ^2 \zeta (3) \left(2 g_k^2-g_{k+1}^2-g_{k-1}^2\right) g_k^2-225 \zeta (5) \left(6 g_k^4-2 \left(g_{k+1}^2+g_{k-1}^2\right) g_k^2-g_{k+1}^4-g_{k-1}^4\right)\\&+26 \pi ^6 g_k^4\Big]+\mathcal{O}(g^{10})\,,
\end{split}
\eeq
in the weak coupling. Plugging \eqref{eq:strongcouplingW} and \eqref{eq:strongcouplinglimitBk} into \eqref{eq:SasfunctionoflogWandB}, we get
\beq
S_k\sim \frac{2\pi}{3}\mu_k=\frac{4\pi}{3}\sqrt{\frac{rg_1^2\cdots g_r^2}{\sum_{i=1}^r\prod_{k\neq i}g_k^2}}\,.
\eeq
up to constant and logarithmic contributions for the strong coupling  limit, with $\mu_k$ taken from \eqref{eq:muinthestrongcoupling}.

\section{Universality of the coupling substitution rule}
\label{universality}

In this section we wish to argue that the effective couplings $f_k(g^2_i)$ that we have been calculating are universal, {\it i.e.}~they are the same for any observable in the $SU(2,1|2)$ sector, up to some scheme dependence that is related to the way the theory is regulated in the infrared.
Firstly, it is important to recall that in  \cite{Pomoni:2013poa,Mitev:2014yba} we argued that in perturbation theory the functions $f_k(g^2_i)$ compute the finite renormalization of the  $\calN=2$ gluon propagator relative  to the  $\calN=4$ one,
\beq
\label{Z-Z}
f_k(g_i^2)=g_k^2+g_k^2\left[\left(\calZ_{g_k}^{\calN=2}\right)^2-\left(\calZ_{g_k}^{\calN=4}\right)^2\right] \, .
\eeq
We checked this proposal by a three loop calculation of the difference between the $\calN=2$ and the $\calN=4$ gluon propagator. The Feynman diagram result, originally done in \cite{Mitev:2014yba}, is presented in \eqref{4loopPropagator}.

In sections \ref{sec:saddlepoint} and \ref{sec:bremsstrahlung}, we  found that up to three loops\footnote{Three loops for  $f(g_i^2)$ is four loops for the observables, the Wilson loop, the Bremsstrahlung function and the entanglement entropy. An insertion of a tree level propagator creates an one loop correction for them and so on.} in the weak coupling expansion and at leading order in the strong coupling the effective couplings are universal, $i.e.$ the  same for the different observables that we studied, namely the Wilson loop, the Bremsstrahlung function and the entanglement entropy. Hence\footnote{One can also derive an effective  coupling $f_S$ for the entanglement entropy. Due to the definition \eqref{eq:SasfunctionoflogWandB}, $f_S$ can be recovered from $f_B$ and $f_W$. Hence, it is enough to discuss the other effective couplings.} , up to that order we see $f_W =f_B =f_S=f$ with
\beq
f_1(g_i^2) = \left\{\begin{array}{ll}g_1^2 + 2\left(g_2^2-g_1^2\right) \left[6\zeta(3) g_1^4 + 20\zeta(5) g_1^4\left(g_2^2+3 g_1^2\right) + \cdots  \right]\ , &  g_1,g_2\rightarrow0\\   2 \frac{g_1^2 g_2^2}{g_1^2 + g_2^2} + \cdots\ , & g_1,g_2\rightarrow \infty\end{array}\right.  \, ,
\eeq
for the $\mathbb{Z}_2$ quiver theory.
These results are identical with the ones in \cite{Mitev:2014yba} and thus in perturbation theory up to three loops the effective couplings are universal, and  compute the relative finite renormalization of the gluon propagators
\eqref{Z-Z}. In the strong coupling the leading term matches the prediction of AdS/CFT \eqref{AdSdictionaty}, \eqref{AdSresult}.

Starting at four loops, we found, see equation \eqref{BWdiscrepancy}, that there are discrepancies between the different effective couplings for the different observables
that  are always proportional to  $\zeta(2)=\frac{\pi^2}{6}$. The first few terms read
\beq
\label{eq:Deltaf}
\Delta f(g_i^2) = 80\zeta(2) g_1^8 (g_1^2-g_2^2)\zeta (5)- g_1^8 (g_1^2-g_2^2) \big(192 g_1^2  \zeta(2)^2 \zeta (5)+ 112(11 g_1^2+5 g_2^2) \zeta(2)\zeta (7)\big)+\cdots \, .
\eeq
Moreover, for the two observables $B$ and $\langle W\rangle$, the difference between  the two effective couplings is due to the dependence in $b$, 
\beq
f_{B} -   f_{W} = \Delta f \sim \frac{\partial f}{\partial b} \, ,
\eeq
see also equation \eqref{BWdiscrepancy}. 
Beginning with this observation and stressing the fact that the parameter $b$
determines how we cut off the low energy momenta for a given calculation, as it is related to the size and shape of the ellipsoid, we wish to argue that the way we extract $f_k(g^2_i)$ suffers from scheme dependence originating in the way the theory is regulated in the infrared.

At the order $g^{10}$, where the discrepancies appear for the first time, the  effective coupling reads
\beq
\label{eq:fresultZ2}
\begin{split}
f_{W;1}(g_1,g_2 )=\cdots&+  2\left(g_2^2-g_1^2\right) g_1^4\Big[ 70\zeta(7) \Big(g_2^4+5g_1^2 g_2^2+8g_1^4\Big)\\&-40\zeta(2)\zeta(5) g_1^4 -2(6\zeta(3))^2  \Big(g_2^4-g_1^2 g_2^2+2 g_1^4 \Big)  \Big]\,.
\end{split}
\eeq
While  it is very clear how to get from Feynman diagrams  the $\zeta(7)$ and the $\zeta(3)^2$ pieces  \cite{Mitev:2014yba}, it is {\it not possible in flat space and  for the massless and finite theories that we are considering to produce a $\zeta(2) \zeta(5)$ term} by one or more Feynman diagrams. This can be understood by carefully looking at the classification of the massless four loop integrals \cite{Baikov:2010hf}. The reader needs to keep in mind that the theories we are looking at are finite \cite{Howe:1983wj} and hence the poles always have to cancel. It would be very important to demonstrate this statement with an explicit calculation, but we leave this for future work.

Let us further stress that the way we have been computing the different $f_k(g_i^2)$ is through localization, which is always done {\it on a sphere \cite{Pestun:2007rz} or on an ellipsoid} \cite{Hama:2012bg}. For those geometries, some of the fields couple conformally to the curvature {\it acquiring an effective mass term} $m^2 = \mu_{R}^2 \propto R \propto (\ell \tilde{\ell})^{-1}$, proportional to a scale set by the Ricci scalar $R$ of the ellipsoid. 
Hence, if we want to reproduce the  $\zeta(2) \zeta(5)$ term by Feynman diagrams, we would have to take into account that {\it some propagators in the integrals that we are computing become massive}.
These mass terms will then be renormalized. For generic theories, this conformal coupling to the curvature usually begins to renormalize starting at two loops. For theories with supersymmetry, the conformal coupling to the curvature will start to renormalize one loop later, at three loops, and we believe that in our case with $\calN=2$ superconformal symmetry the effective mass term will start to renormalize at four loops. Moreover, the presence of massive modes in the loops forces us to specify a {\it mass renormalization scheme}. In the localization results, the  size of the sphere or of the ellipsoid is a parameter independent of $g$. The mass term $m^2 = \mu_{R}^2 \propto R$ does not renormalize, but is instead kept fixed. This is a very particular scheme choice.

\begin{figure}[t]
 \centering
  \includegraphics[height=1.4cm]{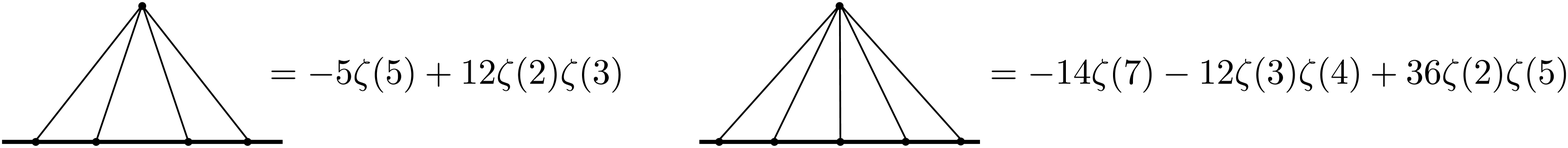}
  \caption{\it 
  The $\zeta(2)$s that we are looking for can be created by massive on-shell propagator diagrams.
  The thick lines indicate the massive propagators, while the thin lines stand for massless ones. The computation is done with the external massive particle momentum  being on the mass shell. }
  \label{fig:Integrals}
\end{figure}

Our next step is to recall examples in 4D QFTs where terms proportional to $\zeta(2)$ are created.  One instance that immediately comes to mind is the computation of bubbles with no external legs. A famous example is the Casimir effect, for which
$\langle T_{00}\rangle = - \frac{\pi^2}{90 L^4} = - \zeta(2)  \frac{1}{15 L^4}$. A second common way to create $\zeta(2)$s is through mass renormalization,  when the mass counterterm is inserted in a bigger diagram and in particular in the large mass expansion \cite{Collins:1984xc,Grozin:2003ak}.
Feynman diagrams like the ones depicted in figure \ref{fig:Integrals} with some on shell propagators are known to create $\zeta(2) \zeta(3)$ or $\zeta(2) \zeta(5)$ terms.\footnote{We thank Erik Panzer for communicating to us the result of the integral on the right hand side of figure \ref{fig:Integrals}.} 
On the left hand side of figure~\ref{fig:Integrals}, taken from \cite{Grozin:2003ak}, a $\zeta(2) \zeta(3)$ finite term is created. The $\zeta(2) \zeta(5)$ term that we are looking for is going to come from a diagram like the one depicted on the right side of figure \ref{fig:Integrals}, at one loop higher than the $\zeta(2) \zeta(3)$ one.

Now that we have made a very particular choice for the mass renormalization scheme, we also have to UV regularize. Let's consider a UV cut off regularization with a UV scale $\Lambda_{\text{UV}}$. From an effective field theory point of view\footnote{We have in mind 
the well known relation between heavy quark QCD and the HQEF \cite{Manohar:2000dt}. In particular,
the decoupling theorem or the ramifications thereof, that is described in chapter 8 of \cite{Collins:1984xc}.}, we should be able to perform the calculation either with
 $\Lambda_{\text{UV}} >> \mu_{R}$, or 
with $\Lambda_{\text{UV}} << \mu_{R}$.
In the latter case, we obtain an effective theory in which the massive fields have decoupled. The relation between the coupling constant in the effective theory without the heavy fields (EFT) and the full theory including the massive fields is given by the matching condition 
\beq 
g_{\text{EFT}}^2= z_{\text{match}} \,g^2_{\text{Full}} \, ,
\eeq
 where $z_{\text{match}} $ is a function that depends on the choice of the scheme and the mass of the heavy fields. For us, the matching condition translates to a relation between the finite coupling renormalization factor  $\calZ_{g}^{\calN=2}$ for the theory in the large mass expansion and the full theory
\beq
\calZ_{g\, \text{EFT}}^{\calN=2} = z_{\text{match}}  \, \calZ_{g\, \text{full}}^{\calN=2} \, .
\eeq
See \cite{Collins:1984xc,Manohar:2000dt} for the general ideology.
Now in this theory with very massive fields, it is easy to see how to create the terms with  $\zeta(2) \zeta(5)$. We just need to look again at the figure~\ref{fig:Integrals}. We see that $\zeta(2) \zeta(3)$ terms can be created when we replace the propagator of the massive fields by the on shell one.  It would be very beautiful to understand why there are no $\zeta(2) \zeta(3)$ terms at order $g^8$, or at any order that we have checked - it is most certainly a cancellation due to supersymmetry.  The $\zeta(2) \zeta(5)$ term that we are looking for is expected to come from diagrams like the one depicted in the right of figure~\ref{fig:Integrals}.  These type of calculations can be done using \cite{Chetyrkin:2000yt} and we leave them for future work.

After all the arguments above, we come to the conclusion that the $\zeta(2)$s  stem from the fact that on the sphere or on the ellipsoid some of the fields  have a mass term $m^2 = \mu_{R}^2 \propto R$. As we go up in loops, the mass is going to be renormalized forcing us to make a choice of a scheme.
At the next loop order, this mass renormalization scheme choice  is going to interfere with our $f(g_i)$ that computes the finite coupling  renormalization of the coupling constant. 
This point was also addressed by Lewkowycz and  Maldacena \cite{Lewkowycz:2013laa} and by Fraser in \cite{Fraser:2015xha}. 
Specifically, in  \cite{Lewkowycz:2013laa} it was discussed that the entanglement entropy computed has a finite ambiguity related to the precise procedure for defining the entropy with the additional finite contributions arising due to the conformal coupling of the scalars to the curvature.

Finally, we wish to conclude this section 
by connecting with our work a recent, independent result in favor of the  existence of the universal coupling substitution rule.
In \cite{Grozin:2014hna,Grozin:2015kna}, the cusp anomalous dimension $\Cusp$ of QCD was compared with the one of $\mathcal{N}=4$ SYM and it was found that, upon replacing the coupling constant $g$ by the light-like cusp anomalous dimension $\cusp$ as
\beq
\label{eq:Omegadefinition}
\Cusp\left(\varphi,g^2\right)=\Omega\left(\varphi,\cusp(g^2)\right)\,,
\eeq
the function $\Omega$ is \textit{independent of the choice of the theory}, at least to three loops. All the dependence on the particular theory stands in $\cusp$. 
It is very important to stress that this result is true for any $\varphi$.
In \eqref{eq:Omegadefinition}, both $\Cusp$ and $\cusp$ have to be computed in the same scheme. Expanding $\Omega$ in $\varphi$, we get
\beq
\Omega(\varphi,\cusp(g^2))=\varphi^2 \tilde{B}(\cusp(g^2))+\mathcal{O}(\varphi^3)\,,
\eeq
and hence comparison with \eqref{eq:definitionB} leads to
$B(g^2)=\tilde{B}(\cusp(g^2))$, 
where $\tilde{B}$ is an \textit{universal function}, at least to 3-loops.
Since 
$B_{\calN=4}(g^2)=\tilde{B}(\cusp_{\calN=4}(g^2))$, $B_{\calN=2}(g^2)=\tilde{B}(\cusp_{\calN=2}(g^2))$
and since from \cite{Pomoni:2013poa} we have the effective coupling relation
$\cusp_{\calN=2}(g^2)=\cusp_{\calN=4}(f(g^2))$, 
we can write
\beq
B_{\calN=2}(g^2)=\tilde{B}(\cusp_{\calN=4}\big(f(g^2)\big)=B_{\calN=4}\big(f(g^2)\big)\,.
\eeq
Following \cite{Grozin:2014hna}, this implies that, at least to 3-loops, the effective coupling $f(g^2)$ is universal, {\it i.e.}~valid for $B$ as well as for the anomalous dimensions.

\section{Conclusions and outlook}
\label{sec:conclusions}

In this article, we calculated in the planar limit the vacuum expectation value of supersymmetric Wilson loops on ellipsoids for the $\calN=2$ cyclic superconformal quivers, by using the localization formula of \cite{Hama:2012bg}. We provided explicit results, both in the weak and in the strong coupling limits. By comparing with $\calN=4$ SYM, we obtained the effective couplings $f_W(g_i^2)$ such that  $W_{\calN=4}(f_W(g_i^2))=W_{\calN=2}(g_i^2)$. 
Using \eqref{eq:bremsstrahlungfiol} and \eqref{eq:SasfunctionoflogWandB} as an input, we extracted from the Wilson loops the Bremsstrahlung functions $B$ and the entanglement entropy $S$ and by comparing them with their $\calN=4$ SYM counterparts we obtained the effective couplings $f_B(g_i^2)$  and  $f_S(g_i^2)$.

From our calculations an important lesson emerges, namely that the effective couplings are universal, $i.e.$ the same for the different observables, up to four loops in the weak coupling and for the leading order in the strong coupling. In particular, for the $\hat{A}_1$ or $\mathbb{Z}_2$ quiver theory, we find that the effective coupling of the first gauge group is 
\beq
f_1(g_i^2) = \left\{\begin{array}{ll}g_1^2 + 2\left(g_2^2-g_1^2\right) \left[6\zeta(3) g_1^4 + 20\zeta(5) g_1^4\left(g_2^2+3 g_1^2\right) + \cdots  \right]\ , &  g_1,g_2\rightarrow0\\   2 \frac{g_1^2 g_2^2}{g_1^2 + g_2^2} + \cdots\ , & g_1,g_2\rightarrow \infty\end{array}\right.  \, .
\eeq
This was checked against the Feynman diagrams calculation in the weak coupling \cite{Mitev:2014yba}, here reviewed in equations
\eqref{RelativeFiniteRenorm} and \eqref{4loopPropagator} as well as the strong coupling leading term \eqref{AdSdictionaty} and \eqref{AdSresult}.
While this is not a direct check of the validity of \eqref{eq:bremsstrahlungfiol} and \eqref{eq:SasfunctionoflogWandB}, it shows that they are consistent with the coupling substitution rule.

Starting at five loops there are discrepancies \eqref{eq:Deltaf} between the effective couplings for the different observables that  are always proportional to $\zeta(2)$. The same exact observation was made by Fraser in \cite{Fraser:2015xha} by comparing Wilson loops in different representations. 
From that, in section \ref{universality}, we draw two lessons. First, from the fact that it is not possible in flat space and for the massless and finite theories that we are considering to produce such a $\zeta(2)$ term by one or more Feynman diagrams, we see that the Wilson loop obtained from localization differs from the flat space circular Wilson loop starting at five loops. Second, we argue that the $\zeta(2)$ discrepancies are an artifact of the localization calculations  being performed on the ellipsoid which imposes hard IR regulators and scheme dependence.
Thus, we propose that the effective couplings are universal up to the fact that one needs to properly take into account this scheme dependence. Following \cite{Pomoni:2013poa}, any anomalous dimension in the purely gluonic, $SU(2,1|2)$, sectors of $\calN=2$ superconformal gauge theories can be obtained from the $\calN=4$ results by directly replacing the $\calN=4$ coupling by the effective couplings.

As an application of the substitution rule, we consider the calculation of the cusp anomalous dimension in $\calN=2$ theories. For $\calN=4$ SYM, the light-like cusp anomalous dimension $\cusp$ is given by
\beq
\cusp_{\calN=4}(g^2)=4 g^2-\frac{4 \pi ^2 g^4}{3}+\frac{44 \pi ^4 g^6}{45}-\left(32 \zeta (3)^2+\frac{292 \pi ^6}{315}\right) g^8+O\left(g^{10}\right)
\eeq
for small $g$ and $\cusp_{\calN=4}(g^2)\sim 2g-\frac{3\log (2)}{2\pi}+\cdots$ for large $g$. Hence, inserting \eqref{eq:fkzeroforcyclicquivers} and \eqref{eq:fkinthestrongcoupling}, we obtain the prediction for the $\mathbb{Z}_2$ quiver
\begin{align}
\cusp(g_i^2)=&4 g_1^2-\frac{4\pi ^2}{3}  g_1^4+\Big[24 \zeta (3) \left(-2 g_1^2+g_2^2+g_5^2\right) g_1^4+\frac{44}{45} \pi ^4 g_1^6\Big]-\Big[32 \zeta (3)^2 g_1^8\\&+16 \pi ^2 \zeta (3) \left(-2 g_1^2+g_2^2+g_5^2\right) g_1^6-80 \zeta (5) \left(6 g_1^4-2 \left(g_2^2+g_5^2\right) g_1^2-g_2^4-g_5^4\right) g_1^4+\frac{292}{315} \pi ^6 g_1^8\Big]+\mathcal{O}\left(g^{10}\right)\,\nonumber
\end{align}
in the weak coupling regime and 
\beq
\cusp(g_i^2)\sim 2  \sqrt{\frac{2g_1^2 g_2^2}{g_1^2+ g_2^2}}
\eeq 
in the strong coupling limit.
Using the three-loop result of \cite{Grozin:2014hna} and inserting $\cusp$ into the function $\Omega$ in equation \eqref{eq:Omegadefinition}  provides us with the full $\Gamma_{\text{cusp}}(\varphi)$ function to three loops for $\calN=2$ SCFTs.

There are many interesting questions and problems left for future work. In our mind, the number one priority is to perform an explicit Feynman diagram calculation on the sphere where the scalars acquire a mass and to explicitly find and compute the diagrams responsible for the first $\zeta(2)\zeta(5)$ discrepancies given in \eqref{eq:Deltaf}. 

Wilson  loops  compute  a big part of the data needed to obtain the
high-energy scattering of charged particles  \cite{Korchemsky:1988si,Korchemsky:1992xv}.
Our diagrammatic studies lead us to believe that light-like polygonal Wilson loops stand a good change to obey the substitution rule, perhaps even in theories with less supersymmetry.
Checking whether the substitution rule works by explicit Feynman diagrams calculations is an important direction worth pursuing. In investigating this direction, it will be paramount to use the appropriate superspace formalism.
Moreover, in  $\mathcal{N}=4$  SYM, polygonal shaped Wilson loops are believed to be exactly dual to scattering amplitudes, see \cite{CaronHuot:2010ek}  and references therein for a recent review. This duality is due to the dual superconformal symmetry which is also believed to combine with the usual conformal symmetry to the Yangian of the full $\mathfrak{psu}(2,2|4)$ integrable model.
If  the substitution rule works for light-like polygonal Wilson loops and gives the correct $\mathcal{N}=2$ results, it implies that there exists a  Yangian symmetry acting on them, something worth checking. However, in strong contrast with $\mathcal{N}=4$ SYM, the dual superconformal symmetry seems to break at two-loops\footnote{The dual superconformal symmetry is broken at two loops by terms depending non-trivially on the kinematics, which are suppressed in the Regge limit. Hence, the amplitude/Wilson loop duality that is broken at two-loops should be restored in the Regge limit. } \cite{Leoni:2014fja,Leoni:2015zxa} destroying the amplitude/Wilson loop duality. 
  It would be very important to understand what this means for the integrability of the $\mathcal{N}=2$ SCFTs and to try to come up with ways to bypass this impasse.

In the case of $\mathcal{N}=4$ SYM, the cusp
anomalous dimension can be obtained by studying  a supersymmetric Wilson loops with $L$ local fields inserted at the cusp \cite{Correa:2012hh,Drukker:2012de}.  This setup is described by TBA equations  very similar to the ones of the spectral problem \cite{Correa:2012hh,Drukker:2012de}.  
However, these TBA equations are simpler and
can be recast in terms of a matrix model \cite{Gromov:2012eu,Gromov:2013qga,Sizov:2013joa} with a spectral curve that can be mapped to  the classical string algebraic curve.
For $\mathcal{N}=2$ SCFTs,  it is currently not clear what happens beyond the $SU(2,1|2)$ sector, but it is worth thinking whether it is possible to derive TBA equations for supersymmetric Wilson loops with local fields from the $SU(2,1|2)$ sector inserted at the cusp.

So far, all the statements we have made about the substitution rule apply to operators long enough, i.e. with anomalous dimensions given by the asymptotic Bethe Ansatz, to evade wrapping corrections. From the TBA perspective, it is counterintuitive to expect that the substitution rule will remain valid when wrapping corrections are taken into account. Nevertheless, from the Feynman diagram point of view, it seems very plausible that the diagrammatic argument of \cite{Pomoni:2013poa} does hold in that case. It would be very illuminating to use L\"uscher techniques \cite{Janik:2010kd} to investigate this further.

Another class of observables that is definitely worth studying is the correlation functions of chiral primary operators, studied in particular in \cite{Papadodimas:2009eu, Baggio:2014sna, Baggio:2014ioa, Baggio:2015vxa}.
The naive coupling substitution for them does not work, due to the fact that finite terms from the non-holomorphic part of the $\calN=2$ effective action $\int d^8\theta \calH(\calW,\bar{\calW})$ contribute to the correlation functions. This is not the case for the anomalous dimensions in the $SU(2,1|2)$ sector where effective vertices from $\int d^8\theta \calH(\calW,\bar{\calW})$ cannot contribute \cite{Pomoni:2013poa}.
Using the methods of \cite{Papadodimas:2009eu, Gerchkovitz:2014gta,Baggio:2014sna, Baggio:2014ioa, Baggio:2015vxa,Gomis:2015yaa}, we can compute the exact Zamolodchikov metric, {\it i.e.}~the metric in theory space, in the planar limit and from that recover the correlation functions of chiral primary operators. This is work in progress. The  Zamolodchikov metric is another  very interesting 
 non-BPS observable. Investigating this direction is currently in progress.

While our results apply for the weak coupling of both the cyclic and the linear quivers, such as $\calN=2$ SCQCD, see figure~\ref{fig:QuiverTheories}, they do not apply to the strong coupling of the linear quivers, since the limit  $g_\nv\rightarrow 0$ does not commute with the strong coupling limit considered in section \ref{subsec:strongcouplingresults}. As already discussed in \cite{Gadde:2009dj}, the strong coupling limit of $\calN=2$ SCQCD is quite subtle. It would be important to understand more about the strong coupling limit of the linear quiver theories, in particular so as to improve our knowledge of their  string duals.

Another theory in which a similar coupling substitution rule applies is ABJM, with the interpolating function $h(\lambda)$ also appearing in the magnon dispersion relation and being computed by comparing with localization techniques \cite{Gromov:2014eha}.
It would be very interesting to study  the Kaluza-Klein reduction of $\mathcal{N}=4$ in the spirit of section 2 of \cite{Alday:2007mf}, to see whether the  interpolating function $h(\lambda)$ can be understood diagrammatically in a spirit similar to ours.

Last but not least, as we discussed in section \ref{universality}, in \cite{Grozin:2014hna,Grozin:2015kna} another, ``experimental'' coupling substitution rule was discovered, in which the coupling $g_{\text{YM}}$ was replaced by  the light-like cusp anomalous dimension $\cusp$ and it was then found that, when so expressed, the full $\Cusp(\varphi)$ is independent of the specific particle content of the gauge theory, at least up to three loops. 
It would be very interesting to try to understand this fact using a diagrammatic argument of the form of \cite{Pomoni:2013poa}, and to try to decide whether there should be other observables that could be obtained in similar ways.

\section*{Note added}
As we have been finishing writing up this note, a closely related paper \cite{Fiol:2015mrp} appeared in the arXiv.

\section*{Acknowledgements}

We thank James Drummond, Bartomeu Fiol, Johannes Henn, Zohar Komargodski, Sven Moch, Volker Schomerus, Matthias Steinhauser,  J\"org Teschner, Konstantin Zarembo for useful discussions and feedback. 
We are grateful to Erik Panzer for sending us the integral on the right hand side of figure \ref{fig:Integrals}.

\appendix

\section{Chebyshev polynomials}
\label{app:chebyshev}

In this appendix, we summarize a couple of useful formulae involving the Chebyshev polynomials $T_l$ and $U_l$ of the first and second kind respectively.
Important for us are the integral identities 
\beq
\label{eq:intTU}
\fint_{-1}^1\sqrt{1-y^2}\frac{U_k(y)}{x-y}dy=\pi T_{k+1}(x), \qquad \fint_{-1}^1\frac{1}{\sqrt{1-y^2}}\frac{T_k(y)}{x-y}dy=-\pi U_{k-1}(x)\,.
\eeq
which are only valid if $x\in (-1,1)$.  The second part of equation \eqref{eq:intTU} implies in particular for $k=1$
\beq
\label{eq:simpleintegralidentities}
\fint_{-\mu_k}^{\mu_k}\frac{dx}{\sqrt{\mu_k^2-x^2}}\frac{1}{z-x}x=-\pi\,.
\eeq
We can generalize \eqref{eq:intTU} for arbitrary $x$ to
to
\beq
\begin{split}
\label{eq:intTU2}
&\frac{1}{\pi}\fint_{-\mu}^{\mu}\sqrt{\mu^2-y^2}\frac{U_k\left(\frac{y}{\mu}\right)}{x-y}dy=\mu T_{k+1}\left(\frac{x}{\mu}\right)-\text{sgn}(x)\Theta(|x|-\mu)\sqrt{x^2-\mu^2}U_k\left(\frac{x}{\mu}\right)\,, \\ &\frac{1}{\pi}\fint_{-\mu}^{\mu}\frac{1}{\sqrt{\mu^2-y^2}}\frac{T_k\left(\frac{y}{\mu}\right)}{x-y}dy=- \frac{1}{\mu} U_{k-1}\left(\frac{x}{\mu}\right)+ \text{sgn}(x)\Theta(|x|-\mu)\frac{T_k\left(\frac{x}{\mu}\right)}{\sqrt{x^2-\mu^2}}\, ,
\end{split}
\eeq
where $\Theta(x)=1$ if $x\geq 0$ and is zero otherwise is the Heaviside function.
For $\mu>0$, one can prove for $x_1\neq x_2$
\beq
\begin{split}
\label{eq:funnyintgen2}
&\frac{1}{\pi}\fint_{-\mu}^{\mu}\frac{dy}{\sqrt{\mu^2-y^2}}\frac{1}{(x_1-y)(x_2-y)}\\&\quad=\frac{\pi \delta(x_1-x_2)\Theta(\mu-|x_1|)}{\sqrt{\mu^2-x_1^2}}+\frac{1}{x_1-x_2}\left(\frac{\text{sgn}(x_2)\Theta(|x_2|-\mu)}{\sqrt{x_2^2-\mu^2}}-\frac{\text{sgn}(x_1)\Theta(|x_1|-\mu)}{\sqrt{x_1^2-\mu^2}}\right)\,.
\end{split}
\eeq
The identity \eqref{eq:funnyintgen2} follows from \eqref{eq:intTU2} as well as from the observation: the equations \eqref{eq:intTU} imply that for a function $\rho(x)$, that we can under some assumptions expand as 
$\rho(x)=\sqrt{\mu^2-x^2}\sum_{n=0}^{\infty}c_n U_n\left(\frac{x}{\mu}\right)$,
we can invert the finite Hilbert kernel and write a $\delta$-function relation like
\beq
\label{eq:invertHilbertkernel}
\rho(x)=-\frac{\sqrt{\mu^2-x^2}}{\pi^2}\fint_{-\mu}^{\mu}\frac{dy}{\sqrt{\mu^2-y^2}}\frac{1}{x-y}\fint_{-\mu}^{\mu}dz\frac{\rho(z)}{y-z}\,.
\eeq

\section{Rewriting the partition functions}
\label{app:rewritingthepartitionfunction}

We refer the reader to appendix C of \cite{Mitev:2014isa} for a definition of the Barnes $\Gamma_2$ function, as well as the function $\Upsilon(x;b)$. Here, we need the product formula for $\Gamma_2$. 
For that, we set for $\Re(s)>2$
\beq
\chi(s;\epsilon_1,\epsilon_2)\colonequals \sum'_{n_1,n_2\geq 0}\frac{1}{(\epsilon_1n_1+\epsilon_2n_2)^s}\,,
\eeq
where the prime removes the value $(n_1,n_2)=(0,0)$ from the sum. The function $\chi(s;\epsilon_1,\epsilon_2)$ can be analytically continued for all $s\in \mathbb{C}$ except for $s=1$ and $s=2$ where there are poles.
 We have the residues
\beq
\text{Res}(\chi(s;\epsilon_1,\epsilon_2),s=1)=\frac{1}{2}\left(\frac{1}{\epsilon_1}+\frac{1}{\epsilon_2}\right),\qquad \text{Res}(\chi(s;\epsilon_1,\epsilon_2),s=2)=\frac{1}{\epsilon_1\epsilon_2}
\eeq
and the finite parts
\beqa
\text{Res}\Big(\frac{\chi(s;\epsilon_1,\epsilon_2)}{s-1},s=1\Big)&=&-\frac{\log \epsilon_1}{\epsilon_1}+\frac{1}{2}\left(\frac{1}{\epsilon_1}-\frac{1}{\epsilon_2}\right)\log \epsilon_2+\frac{\gamma}{\epsilon_1}+\frac{\gamma}{2\epsilon_2}-\frac{1}{2\epsilon_1}\log2\pi\nonumber\\&&-\frac{i}{b}\int_{0}^{\infty}\frac{\psi(i\frac{\epsilon_1}{\epsilon_2}y+1)-\psi(-i\frac{\epsilon_1}{\epsilon_2}y+1)}{e^{2\pi y}-1}dy\nonumber\\
\text{Res}\Big(\frac{\chi(s;\epsilon_1,\epsilon_2)}{s-2},s=2\Big)&=&\frac{\zeta(2) }{\epsilon_1^2}+\frac{\zeta(2)}{2\epsilon_2^2}+\frac{1}{\epsilon_1\epsilon_2}\left(\gamma-1-\log \epsilon_2\right)\nonumber\\&&-\frac{i}{\epsilon_2}\int_{0}^{\infty}\frac{\zeta_H(2,i\frac{\epsilon_1}{\epsilon_2}y+1)-\zeta_H(2,-i\frac{\epsilon_1}{\epsilon_2}y+1)}{e^{2\pi y}-1}dy\,,
\eeqa
where $\psi$ is the digamma function, $\gamma$ is the Euler - Mascheroni constant and $\zeta_H(s,q)$ is the Hurwitz-$\zeta$ function with ($\Re(s)>1$ and $\Re(q)>0$)
$\zeta_H(s,q)\colonequals \sum_{n=0}^{\infty}\frac{1}{(q+n)^s}$.
Finally, using the shorthands 
\beq
\label{eq:resaandresb}
\resa\colonequals \text{Res}(\frac{\chi(s;\epsilon_1,\epsilon_2)}{s-1},s=1)\,,\qquad \resb\colonequals \text{Res}(\frac{\chi(s;\epsilon_1,\epsilon_2)}{s-2},s=2)+\text{Res}(\chi(s;\epsilon_1,\epsilon_2),s=2)\,,
\eeq
we present the product formula for $\Gamma_2$
\beq
\label{eq:defGamma2product}
\Gamma_2(x;\epsilon_1,\epsilon_2)=\frac{e^{-\resa x +\frac{\resb x^2}{2}}}{x} \prod'_{n_1,n_2\geq 0}\frac{e^{\frac{x}{\epsilon_1n_1+\epsilon_2n_2}-\frac{x^2}{2(\epsilon_1n_1+\epsilon_2n_2)^2}}}{1+\frac{x}{\epsilon_1n_1+\epsilon_2n_2}}\,.
\eeq
Let us introduce two new special functions that we need for \eqref{eq:1loopHosomichiVersion2}. Specifically, we define
\beq
\label{eq:definitionHvandHh}
\begin{split}
H_v(x;b)\colonequals& \prod_{m,n=0}^{\infty}\sqrt{\left(1+\frac{x^2}{(b(m+1)+b^{-1}n)^2}\right)\left(1+\frac{x^2}{(bm+b^{-1}(n+1))^2}\right)}\nonumber\\&\times e^{-\frac{x^2}{2(b(m+1)+b^{-1}n)^2}-\frac{x^2}{2(bm+b^{-1}(n+1))^2}}\,,\\
H_h(x;b)\colonequals& e^{-\zeta(2)\frac{b^2+b^{-2}}{2}x^2}\prod_{m,n=0}^{\infty}\left(1+\frac{x^2}{\big(b(m+\nicefrac{1}{2})+b^{-1}(n+\nicefrac{1}{2})\big)^2}\right)e^{-\frac{x^2}{(b(m+1)+b^{-1}(n+1))^2}}\,.
\end{split}
\eeq
It is easy to see that both these functions are even in $x$, that $H_v(0;b)=H_h(0;b)=1$  and that
\beq
\lim_{b\rightarrow 1}H_v(x;b)=\lim_{b\rightarrow 1}H_h(x;b)=H(x)=\prod_{n=1}^n\left(1+\frac{x^2}{n^2}\right)^ne^{-\frac{x^2}{n}}\,.
\eeq
Furthermore, $H_a(x;b)=H_a(x;b^{-1})$,  both for $a=v$ and for $a=h$. We also need to consider the logarithms of the functions $H_v$ and $H_h$ defined in \eqref{eq:definitionHvandHh}. Let us define
\beq
\label{eq:definitionsKvandKh}
K_v(x;b)\colonequals -\frac{d}{dx}\log(H_v(x;b))\,,\qquad K_h(x;b)\colonequals-\log \frac{d}{dx}\log(H_h(x;b))\,.
\eeq
We can easily compute the logarithms of $H_v$ and $H_h$, finding that 
\beq
\label{eq:KvandKhpowerseries}
\begin{split}
K_v(x;b)&=-2\sum_{n=1}^{\infty}(-1)^{n}x^{2n+1}\frac{\zeta_2(b,2n+2;b,b^{-1})+\zeta_2(b^{-1},2n+2;b,b^{-1})}{2}\,,\\
K_h(x;b)&=-2\sum_{n=1}^{\infty}(-1)^{n}x^{2n+1}\zeta_2(\nicefrac{(b+b^{-1})}{2},2n+2;b,b^{-1})\,,
\end{split}
\eeq
where we have used  Barnes $\zeta_2$ function
$\zeta_2(x,s;\epsilon_1,\epsilon_2)\colonequals \sum_{n_1,n_2\geq 0}\frac{1}{(x+n_1\epsilon_1+n_2\epsilon_2)^s}$, 
convergent for $\text{Re}(s)\geq 2$.
We need the following special values 
\begin{align}
&\zeta_2(1,2n+2;1,1)=\zeta(2n+1)\,,&
&\frac{d}{db}\zeta_2(b,2n+2;b,b^{-1})_{\big| b=1}=-(2n+2)\zeta(2n+2)\,,&\nonumber\\
&\frac{d}{db}\zeta_2(b^{-1},2n+2;b,b^{-1})_{\big| b=1}=(2n+2)\zeta(2n+2)\,,&
&\frac{d}{db}\zeta_2(\nicefrac{(b+b^{-1})}{2},2n+2;b,b^{-1})_{\big| b=1}=0\,.&
\end{align}
Hence $K_v(x;1)=K_h(x;1)=K(x)\colonequals-2\sum_{n=1}^{\infty}(-1)^{n}\zeta(2n+1)x^{2n+1}$ and 
\beq
\label{eq:derivativesKvandKh}
\frac{d}{db}K_v(x;b)_{|b=1}=\frac{d}{db}K_h(x;b)_{|b=1}=0\,.
\eeq
Similarly, expanding in $b$ around $b=1$, replacing the summation variables as $n_1=\nicefrac{(r+s)}{2},$ $n_2=\nicefrac{(r-s)}{2}$ with $r\in \mathbb{N}_0$ and $s\in\{-r,-r+2,\ldots, r\}$, we find the expansions
\beq
\begin{split}
K_v(x;b)=&-2\sum_{n=1}^{\infty}(-1)^{n}x^{2n+1}\Big[\zeta(2n+1)+\frac{4}{3}(n+1)\big(n\zeta(2n+1)+(2n+3)\zeta(2n+3)\big)(b-1)^2\\&-4 (n+1) \big(n \zeta (2 n+1)+(2 n+3) \zeta (2 n+3)\big)(b-1)^3+\mathcal{O}(b-1)^4\Big]\,,\\
K_h(x;b)=&-2\sum_{n=1}^{\infty}(-1)^{n}x^{2n+1}\Big[\zeta(2n+1)+\frac{2}{3} (n+1) \big(2 n \zeta (2 n+1)-(2 n+3) \zeta (2 n+3)\big)(b-1)^2\\&-2 (n+1) \big(2 n \zeta (2 n+1)-(2 n+3) \zeta (2 n+3)\big)(b-1)^3+\mathcal{O}(b-1)^4\Big]\,,
\end{split}
\eeq
where we have used ($B_k(1)$  is the value of the $k^{\text{th}}$ Bernoulli polynomial at 1)
\beq
\sum_{\substack{s=-r\\\text{step }2}}^rs^m=\delta _{(m \bmod 2),0} \left(\delta _{m,0}+\frac{2^{m+1}}{m+1} \sum _{k=0}^m B_k(1) \binom{m+1}{k} \left(\frac{r}{2}\right)^{m+1-k}\right)\,.
\eeq

Let us now return to the partition functions on the ellipsoid.
For conformal field theories, we can claim that we can rewrite \eqref{eq:partitionfunctionHosomichi} as \eqref{eq:partitionfunctionHosomichi2}
with $\parf^{\text{vect}}_{\text{1-loop}}$ and $\parf^{\text{hyper}}_{\text{1-loop}}$ given in \eqref{eq:1loopHosomichiVersion2}.

\proof We rescale the integration variables in \eqref{eq:partitionfunctionHosomichi} by $\sqrt{\ell\tilde{\ell}}$. This does not change the results for conformal field theories. We now consider the partition functions \eqref{eq:1loopHosomichi} separately. First, we look at the vector multiplet contributions. We use the definitions of appendix C of \cite{Mitev:2014isa} to write
\beq
\label{eq:splitUpsilonVandermonde}
\Upsilon(x;b)=-x b^{(b-b^{-1})x}\frac{\Gamma(-bx)\Gamma(-b^{-1}x)\Gamma_2(\nicefrac{Q}{2};b,b^{-1})^2}{2\pi \Gamma_2(x;b,b^{-1})\Gamma_2(-x;b,b^{-1})}\,,
\eeq
in order to split away the Vandermonde determinant contribution. We get
\beq
\begin{split}
\parf^{\text{vect}}_{\text{1-loop}}&=\prod_{i<j=1}^N\Upsilon\big(i(a_i-a_j);b\big)\Upsilon\big(-i(a_i-a_j);b\big)\\
&=\prod_{i<j=1}^N(a_i-a_j)^2\frac{\prod_{s=\pm}\left[\Gamma(sib(a_i-a_j))\Gamma(sib^{-1}(a_i-a_j))\right]\Gamma_2(\nicefrac{Q}{2})^4}{(2\pi)^2 \Gamma_2(i(a_i-a_j))^2\Gamma_2(-i(a_i-a_j))^2}
\end{split}
\eeq
Since we are only interested in the computation of the Wilson loops \eqref{eq:WilsonloopHosomichi}, we should be able to rescale the partition function even by an $b$-dependent function, so that we can drop the $\nicefrac{\Gamma_2(\nicefrac{Q}{2})^4}{(2\pi)^2}$ part. Hence, we can use instead
\beq
\label{eq:calZtemporarysplit}
\parf^{\text{vect}}_{\text{1-loop}}=\prod_{i<j=1}^N(a_i-a_j)^2\prod_{i<j=1}^N\frac{\Gamma(ib(a_i-a_j))\Gamma(-ib(a_i-a_j))}{\Gamma_2(i(a_i-a_j))\Gamma_2(-i(a_i-a_j))}\prod_{i<j=1}^N(b\leftrightarrow b^{-1})
\eeq
Using \eqref{eq:defGamma2product} and $\Gamma(x)=\frac{e^{-\gamma x}}{x}\prod_{n=1}^{\infty}\frac{e^{\frac{x}{n}}}{1+\frac{x}{n}}$, we find
\beq
\begin{split}
\frac{\Gamma(ibx)\Gamma(-ibx)}{\Gamma_2(ix)\Gamma_2(-ix)}&=\frac{e^{\resb x^2}}{b^2}\frac{\prod_{n=1}^{\infty}\left(1+\frac{x^2}{(b^{-1}n)^2}\right)}{\prod_{m,n\geq 0}'\frac{e^{\frac{x^2}{(bm+b^{-1}n)^2}}}{1+\frac{x^2}{(bm+b^{-1}n)^2}}}\,,
\end{split}
\eeq
where $\resb$ was defined in \eqref{eq:resaandresb}.
Hence it follows that we can write \eqref{eq:calZtemporarysplit} as
\beq
\label{eq:vectorver2}
\parf^{\text{vect}}_{\text{1-loop}}=\prod_{i<j=1}^N(a_i-a_j)^2\prod_{i,j=1}^N\prod_{i<j=1}^Ne^{\big(\resb-\frac{b^2+b^{-2}}{2}\zeta(2)\big)(a_i-a_j)^2}H_v(a_i-a_j;b)\,.
\eeq
Second, let us look at the hyper multiplet contribution. We write down explicitly
\beq
\Upsilon(ix+\nicefrac{Q}{2})=\frac{\Gamma_2(\nicefrac{Q}{2})^2}{\Gamma_2(\nicefrac{Q}{2}+ix)\Gamma_2(\nicefrac{Q}{2}-ix)}
\eeq
and use \eqref{eq:defGamma2product} to get
\beq
\begin{split}
\Upsilon(ix+\nicefrac{Q}{2})&=e^{x^2\resb}\frac{(\nicefrac{Q}{2}+ix)(\nicefrac{Q}{2}-ix)}{\big(\nicefrac{Q}{2}\big)^2} \prod_{m,n\geq 0}'e^{-\frac{x^2}{(bm+b^{-1}n)^2}}\frac{\left(1+\frac{\nicefrac{Q}{2}+ix}{bm+b^{-1}n}\right)\left(1+\frac{\nicefrac{Q}{2}-ix}{bm+b^{-1}n}\right)}{\left(1+\frac{\nicefrac{Q}{2}}{bm+b^{-1}n}\right)^2}\\
&=e^{x^2\resb}\left(1+\frac{4x^2}{Q^2}\right)\prod_{m,n\geq 0}'e^{-\frac{x^2}{(bm+b^{-1}n)^2}}\left(1+\frac{x^2}{\left(b\big(m+\nicefrac{1}{2})+b^{-1}\big(n+\nicefrac{1}{2}\big)\right)^2}\right)\,.
\end{split}
\eeq
At this point, we can split the product for the exponential pieces as 
\beq
\prod_{m,n=0}^{\infty}(\cdots)=\prod_{m,n=1}^{\infty}(\cdots)\prod_{m=1}^{\infty}(\cdots)\prod_{n=1}^{\infty}(\cdots)\,,
\eeq
 express the $\prod_{m=1}^{\infty}(\cdots)\prod_{n=1}^{\infty}(\cdots)$ piece as $e^{-x^2(b^2+b^{-2)}\zeta(2)}$ and absorb the factor $\left(1+\frac{4x^2}{Q^2}\right)$ inside the remaining product to obtain
\beq
\begin{split}
\label{eq:UpsilonxplushalfQ}
\Upsilon(ix+\nicefrac{Q}{2})&=e^{x^2(2\resb-(b^2+b^{-2})\zeta(2))} \prod_{m,n=0}^{\infty}\left(1+\frac{x^2}{\big(b(m+\nicefrac{1}{2})+b^{-1}(n+\nicefrac{1}{2})\big)^2}\right)e^{-\frac{x^2}{(b(m+1)+b^{-1}(n+1))^2}}\\
&=e^{x^2\big(\resb-\frac{b^2+b^{-2}}{2}\zeta(2)\big)}H_h(x;b)\,.
\end{split}
\eeq
Hence, feeding \eqref{eq:UpsilonxplushalfQ} into \eqref{eq:1loopHosomichi} leads to
\beq
\label{eq:hyperver2}
\begin{split}
\parf^{\text{hyper}}_{\text{1-loop}}&=\prod_{i,j=1}^N\Upsilon\big(i(a_i^{(1)}-a_j^{(2)})+\nicefrac{Q}{2};b\big)^{-1}=\prod_{i,j=1}^Ne^{-(a_i^{(1)}-a_j^{(2)})^2\big(\resb-\frac{b^2+b^{-2}}{2}\big)}H_h(a_i^{(1)}-a_j^{(2)};b)^{-1}\,.
\end{split}
\eeq
Putting \eqref{eq:vectorver2} and \eqref{eq:hyperver2} together, the exponential terms cancel for conformal field theories, leaving us with the desired result \eqref{eq:1loopHosomichiVersion2}.

\qed

\section{The weak coupling expansion}
\label{app:weakcouplingexpansion}

In this appendix, we wish to take the set of linear integral equations \eqref{eq:saddlepoint3} and find an approximate solution for small values of the couplings. Our computations follow the principles outlined in \cite{Passerini:2011fe}. For our purposes, we fix an integer $\app\geq 1$ and expand the kernels $K_v$ and $K_h$ as
\beq
\label{eq:expandK}
K_a(x)\approx-2\sum_{n=1}^\app(-1)^n\coef_a(n)x^{2n+1}\,,
\eeq
where $a\in\{v,\,h\}$, the coefficients $\coef_a(n)$ can be extracted from equation \eqref{eq:KvandKhpowerseries} and we have suppressed the $b$ dependence. This expansion is sufficient in order to obtain the results up to order $g^{2(P+1)}$ in the couplings.
We  have for a given eigenvalue density $\rho_k$ the expression
\beq
\label{eq:expandKintegraterho}
\int_{-\mu}^{\mu}dz\rho_k(z)K_a(y-z)=-2\sum_{n=1}^\app(-1)^n\coef_a(n)\sum_{s=0}^n\binom{2n+1}{2s}y^{2(n-s)+1}\moment_{2s}^{(k)}\,,
\eeq
where $\moment_i^{(k)}$ is the $i$-th moment of the density $\rho_k$, i.e.
\beq
\label{eq:definitionmoment}
\int_{-\mu}^{\mu}\rho_k(x)x^i=\moment_i^{(k)}\,.
\eeq
Observe that the odd moments have to vanish due to the symmetry of the densities.
Using Chebyshev polynomials, we can derive the integral formula
\beq
\label{eq:principalvalueintegral1}
 \fint_{-\mu}^{\mu}\frac{dy}{x-y}\frac{y^{n}}{\sqrt{\mu^2-y^2}}=-\pi\sum_{t=0}^{\lfloor \frac{n-1}{2}\rfloor}\frac{(t+1)C_t}{4^t}\mu^{2r}x^{n-1-2t}\,.
\eeq
where $C_r=\frac{1}{r+1}\binom{2r}{r}$ is the $r$-th Catalan number. Plugging \eqref{eq:expandKintegraterho} and \eqref{eq:principalvalueintegral1} into \eqref{eq:saddlepoint3}, we the following result for the $k^{\text{th}}$ density:
\beq
\begin{split}
\label{eq:rhokasmoments}
\rho_k(x)=&\frac{1}{2\pi g_k^2}\sqrt{\mu_k^2-x^2}-\frac{2}{\pi}\sqrt{\mu_k^2-x^2}\sum_{n=1}^\app(-1)^n\sum_{l=1}^\nv\Big[\delta_{kl}\coef_v(n)-\frac{\delta_{k,l+1}+\delta_{k,l-1}}{2}\coef_h(n)\Big]\\&\times \sum_{j=0}^n\binom{2n+1}{2j}\moment_{2j}^{(l)}\sum_{t=0}^{n-j}\frac{(t+1)C_t}{4^t}\mu_k^{2t}x^{2(n-j-t)}\,.
\end{split}
\eeq
We thus have expressed each of the densities $\rho_k$ as functions of its $\app$ first non-trivial moments $\moment_{2i}^{(k)}$, $i=1,\ldots, \app$. Computing the moments by plugging  \eqref{eq:rhokasmoments} into the definition \eqref{eq:definitionmoment}, we obtain a set of $\nv\times \app$ linear equations for the same number of variables $\moment_{2i}^{(k)}$:
\beq
\begin{split}
\label{eq:equationsforthemoments}
&\sum_{j=1}^\app\sum_{l=1}^\nv\Bigg[\delta_{ij}\delta_{kl}+\sum_{n=j}^\app(-1)^n\Big[\delta_{kl}\coef_v(n)-\frac{\delta_{k,l+1}+\delta_{k,l-1}}{2}\coef_h(n)\Big]\binom{2n+1}{2j}\\&\times \frac{\mu_k^{2(n+i-j+1)}}{4^{n+i-j}}\binom{2i}{i}\binom{2(n-j)}{n-j}\frac{2(n-j)+1}{n-j+i+1}
\Bigg]\moment_{2j}^{(l)}=\\&=\frac{1}{g_k^2}\frac{C_i}{4^{i+1}}\mu_k^{2(i+1)}-\sum_{n=1}^\app(-1)^n\Big[\coef_v(n)-\coef_h(n)\Big]\frac{\mu_k^{2(n+i+1)}}{4^{n+i}}\binom{2i}{i}\binom{2n}{n}\frac{2n+1}{n+i+1}\,,
\end{split}
\eeq
for $i=1,\ldots, \app$  and $k=1,\ldots, \nv$. 
In \eqref{eq:equationsforthemoments},  we have used the integral formula
\beq
\int_{-\mu}^{\mu}dx \sqrt{\mu^2-x^2}x^{2s}=\frac{2\pi C_s}{4^{s+1}}\mu^{2(s+1)}\,,
\eeq
the fact that $m_0^{(l)}=1$ $\forall l$ and the following formula for the Catalan numbers
\beq
\sum_{t=0}^m(t+1)C_tC_{m+i-t}=\binom{2i}{i}\binom{2m}{m}\frac{2m+1}{m+i+1}\,.
\eeq
Observe that the last line of \eqref{eq:equationsforthemoments} vanishes for $b= 1$, since in that case $\coef_v(n)=\coef_h(n)$.
The set of linear equations \eqref{eq:equationsforthemoments}  allows us to solve for the moments as functions of the densities widths $\mu_k$ and of the couplings $g_k^2$. The $\mu_k$ are then expressed as functions of the coupling by normalizing the densities. Specifically, plugging \eqref{eq:rhokasmoments} into the normalization condition for the densities \eqref{eq:rhodefandnormalization}, we arrive at 
\begin{align}
\label{eq:normalizationequationforthewidths}
1=&\frac{\mu_k^2}{4g_k^2}-\sum_{l=1}^\nv\sum_{n=1}^\app(-1)^n\Big[\delta_{kl}\coef_v(n)-\frac{\delta_{k,l+1}+\delta_{k,l-1}}{2}\coef_h(n)\Big]\sum_{j=1}^n\binom{2n+1}{2j}\frac{\mu^{2(n-j+1)}_k\moment_{2j}^{(l)}}{4^{n-j}}
\binom{2(n-j)}{n-j}\frac{2(n-j)+1}{n-j+1}\nonumber\\
&-\sum_{n=1}^\app(-1)^n\Big[\coef_v(n)-\coef_h(n)\Big]\frac{\mu_k^{2(n+1)}}{4^n}\binom{2n}{n}\frac{2n+1}{n+1}\,.
\end{align}
Again, the last term of \eqref{eq:normalizationequationforthewidths} vanishes for $b=1$. 
Inserting in \eqref{eq:normalizationequationforthewidths} the expressions obtained from \eqref{eq:equationsforthemoments} for the moments allows us to solve for the $\mu_k$ as functions of the $g^2_k$. 

In order to solve equations \eqref{eq:equationsforthemoments} and \eqref{eq:normalizationequationforthewidths}, it is numerically good to linearize in the couplings as
\beq
\mu_k=2g_k\left(1+\sum_{i=1}^{P+1}\alpha_i^{(k)}g_k^{2i}\right)\,,\qquad \mathfrak{m}_{2i}^{(k)}=C_ig_k^{2i}\left(1+\sum_{j=1}^{P+1-i}\beta_{j,i}^{(k)}g_k^{2i}\right)\,.
\eeq
We then expand the equations in powers of $g_k$ up to the power $2(P+1)$. Once we have solved for the coefficients  $\alpha_i^{(k)}$ and the $\beta_{j,i}^{(k)}$, we have obtained the widths and the moments and the Wilson loop expectation values \eqref{eq:Wilsonpmdefinition} can be expressed as
\beq
\langle W_k^{\pm}\rangle= \sum_{n=0}^{\infty}\frac{(2 \pi )^{2 n}}{(2 n)!}b^{\pm 2n}\moment_{2n}^{(k)}=2\sum_{n=0}^{\infty}\frac{(-1)^{n+1}\zeta(2n)}{B_{2n}}b^{\pm 2n}\moment_{2n}^{(k)}\,,
\eeq
where $B_n$ are the Bernoulli numbers.

For the $\mathbb{Z}_2$ quiver, we obtain the expansion up to order $(b-1)^2$
\beq
\label{eq:resultforWplusbforZ2}
\begin{split}
\langle W_1^{+}\rangle&=1+2 \pi ^2 g_1^2+\frac{4}{3} \pi ^4 g_1^4+\frac{4}{9} \pi ^2 g_1^4 \Big[54 \zeta (3) \left(g_2^2-g_1^2\right)+\pi ^4 g_1^2\Big]\\&+\frac{4}{45} \pi ^2 g_1^4 \Big[360 \pi ^2 \zeta (3) \left(g_2^2-g_1^2\right) g_1^2+900 \zeta (5) (g_1-g_2) (g_1+g_2) \left(3 g_1^2+g_2^2\right)+\pi ^6 g_1^4\Big]\\&
+\frac{8}{675} \pi ^2 g_1^4 \Big[-1350 \pi ^4 \zeta (3) \left(g_1^2-g_2^2\right) g_1^4+2250 \pi ^2 \zeta (5) \left(13 g_1^4-9 g_2^2 g_1^2-4 g_2^4\right) g_1^2\\&
+675 \left(g_1^2-g_2^2\right) \left(36 \zeta (3)^2 \left(2 g_1^4-g_2^2 g_1^2+g_2^4\right)-35 \zeta (7) \left(8 g_1^4+5 g_2^2 g_1^2+g_2^4\right)\right)+\pi ^8 g_1^6\Big]+\cdots\\
&+\Bigg[4 \pi ^2 g_1^2+\frac{16}{3} \pi ^4  g_1^4+\frac{8}{3} \pi ^2  g_1^4 \Big[18 \zeta (3) \left(g_2^2-g_1^2\right)+\pi ^4 g_1^2\Big]\\&
+\frac{32}{45} \pi ^2 g_1^4 \Big[180 \pi ^2 \zeta (3) \left(g_2^2-g_1^2\right) g_1^2+225 \zeta (5) (g_1-g_2) (g_1+g_2) \left(3 g_1^2+g_2^2\right)+\pi ^6 g_1^4\Big]\\
&+\frac{16}{135} \pi ^2 g_1^4 \Big[810 \pi ^4 \zeta (3) \left(g_2^2-g_1^2\right) g_1^4+900 \pi ^2 \zeta (5) \left(13 g_1^4-9 g_2^2 g_1^2-4 g_2^4\right) g_1^2\\&+135 \left(g_1^2-g_2^2\right) \left(36 \zeta (3)^2 \left(2 g_1^4-g_2^2 g_1^2+g_2^4\right)-35 \zeta (7) \left(8 g_1^4+5 g_2^2 g_1^2+g_2^4\right)\right)+\pi ^8 g_1^6\Big]+\cdots \Bigg](b-1)\\
&+\Bigg[
2 \pi ^2 g_1^2+8 \pi ^4 g_1^4+\frac{4}{3} \pi ^2  g_1^4 \Big[66 \zeta (3) \left(g_2^2-g_1^2\right)-120 \zeta (5) \left(4 g_1^2+g_2^2\right)+5 \pi ^4 g_1^2\Big]\\
&+\frac{16}{45} \pi ^2 g_1^4 \Big[780 \pi ^2 \zeta (3) \left(g_2^2-g_1^2\right) g_1^2-75 \zeta (5) \left(\left(34 \pi ^2-81\right) g_1^4+2 \left(27+4 \pi ^2\right) g_2^2 g_1^2+27 g_2^4\right)\\&+1575 \zeta (7) \left(15 g_1^4+4 g_2^2 g_1^2+2 g_2^4\right)+7 \pi ^6 g_1^4\Big]
+\frac{8}{15} \pi ^2 g_1^4 \Big[-10 \pi ^4 g_1^4 \big((53 \zeta (3)+88 \zeta (5)) g_1^2\\
&+(20 \zeta (5)-53 \zeta (3)) g_2^2\big)+140 \pi ^2 g_1^2 \big((65 \zeta (5)+166 \zeta (7)) g_1^4+45 (\zeta (7)-\zeta (5)) g_2^2 g_1^2\\&+20 (\zeta (7)-\zeta (5)) g_2^4\big)+15 \big(8 \left(57 \zeta (3)^2+480 \zeta (5) \zeta (3)-595 \zeta (7)-1806 \zeta (9)\right) g_1^6\\&-3 \left(228 \zeta (3)^2+720 \zeta (5) \zeta (3)-595 \zeta (7)+840 \zeta (9)\right) g_2^2 g_1^4+4 \big(6 \zeta (3) (19 \zeta (3)-40 \zeta (5))+595 \zeta (7)\\&-840 \zeta (9)\big) g_2^4 g_1^2-\left(228 \zeta (3)^2+720 \zeta (5) \zeta (3)-595 \zeta (7)+840 \zeta (9)\right) g_2^6\big)+\pi ^8 g_1^6\Big]+\dots
\Bigg](b-1)^2\\&+\mathcal{O}(b-1)^3
\end{split}
\eeq
From \eqref{eq:resultforWplusbforZ2} one can extract the effective coupling \eqref{eq:effcouplingbforZ2part0}ff.

\section{The strong coupling limit}
\label{app:strongcouplinglimit}

In this subsection, we want to  make a strong coupling analysis of the saddle point equations \eqref{eq:saddlepoint3} in the case in which $b=1$, i.e.~on the sphere. In so doing, we will follow the same procedure as explained in the appendices of \cite{Passerini:2011fe}, with some additional details and complications.
Before we begin in earnest with the study of the strong coupling limit, it is necessary to derive some integral identities. Of particular importance are \eqref{eq:intTU2} and \eqref{eq:funnyintgen2}.
We  define the function
\beq
\theta(x)\colonequals x\coth(\pi x)\,.
\eeq
We use the same conventions for the Fourier transform as \cite{Passerini:2011fe}, i.e.
$\hat{f}(\omega)\colonequals \int_{-\infty}^{\infty}dx e^{ix\omega}f(x)$.
so that in the sense of distributions we get
\beq
\label{eq:fourriertranformoftheta}
\hat{\theta}(\omega)=-\frac{1}{2\sinh^2(\frac{\omega}{2})}\,.
\eeq
From equation (4.14) of \cite{Passerini:2011fe}, we take the following ``formal'' integral formula for $K(x)=K_v(x;1)=K_h(x;1)$
\beq
\label{eq:integralformulaforK1}
K(x)=\fint_{-\infty}^{\infty}\frac{\theta(w)}{x-w}dw\,.
\eeq
To make \eqref{eq:integralformulaforK1} correct, one needs to shift the argument of $K(x)$ by $z$ and average over $z$ such that the first two moments vanish. Specifically, we observe that
\beq
\begin{split}
\int_{-\mu}^{\mu}dy\rho(y)(K(x-y)-K(x))&=\int_{-\mu}^{\mu}dy\rho(y)\left(-yK'(x)+\frac{1}{2}y^2K''(x)+\cdots\right)\\&=-\moment_1K'(x)+\frac{1}{2}\moment_2K''(x)+\cdots =\frac{1}{2}\moment_2K''(x)+\cdots 
\end{split}
\eeq
so that \eqref{eq:integralformulaforK1} is applicable, i.e.
\beq
\label{eq:HilberttransformofK}
\int_{-\mu}^{\mu}dy\rho(y)(K(x-y)-K(x))=\fint_{-\infty}^{\infty}\frac{dw}{x-w}\int_{-\mu}^{\mu}dy\rho(y)\left[\theta(w-y)-\theta(w)\right]\,,
\eeq
which has been checked numerically. If we use \eqref{eq:HilberttransformofK} for the kernel $K$ as well as equation  \eqref{eq:funnyintgen2}, we are able to show\footnote{The ``sgn'' part of the equation is a convention that can be absorbed in the definition of the square roots.} 
\begin{align}
\label{eq:kernelformula}
&\frac{1}{\pi^2}\fint_{-\mu}^{\mu}\frac{dy}{x-y}\sqrt{\frac{\mu^2-x^2}{\mu^2-y^2}}\int_{-\mu}^{\mu}dz \rho(z)\Big[K(y-z)-K(y)\Big]\nonumber\\&=-\frac{\sqrt{\mu^2-x^2}}{\pi^2}\fint_{-\infty}^{\infty}dw\int_{-\mu}^{\mu}dz\rho(z)\big(\theta(w-z)-\theta(w)\big)\fint_{-\mu}^{\mu}\frac{dy}{(x-y)(w-y)}\frac{1}{\sqrt{\mu^2-y^2}}\nonumber\\&=\theta(x)-(\rho\star \theta)(x)+\frac{1}{\pi}\int_{|w|>\mu}\frac{\text{sgn(w)dw}}{w-x}\sqrt{\frac{\mu^2-x^2}{w^2-\mu^2}}\Big[(\rho\star \theta)(w)-\theta(w)\Big],
\end{align}
where we have defined the convolution as
$(\rho\star\theta)(x)\colonequals \int_{-\mu}^{\mu}dy\rho(y)\theta(x-y)$.

Having set up the necessary additional identities, we can start our analysis of the strong coupling behavior of the saddle point equations. We take the system of $\nv$ coupled integral equations for the cyclic quivers \eqref{eq:saddlepoint3} for $b=1$ and using \eqref{eq:invertHilbertkernel} we rewrite them as:
\beq
\rho_k(x)=\frac{\sqrt{\mu_k^2-x^2}}{2\pi g_k^2}-\frac{1}{\pi^2}\int_{-\mu_k}^{\mu_k}\frac{dy}{x-y}\sqrt{\frac{\mu_k^2-x^2}{\mu_k^2-y^2}}\sum_{l=1}^\nv\frac{\car_{kl}}{2}\int_{-\mu_l}^{\mu_l}\rho_l(z)\left[K(y-z)-K(z)\right]dz\,,
\eeq
where $\car_{kl}\colonequals 2\delta_{kl}-\delta_{k,l+1}-\delta_{k,l-1}$ is the $\SU{\nv}$ Cartan matrix and we have used $\sum_{l=1}^\nv\car_{kl}=0$. Using  \eqref{eq:kernelformula}, we arrive at 
\begin{align}
\label{eq:definitionofthedrivingterm}
\rho_k(x)-&\sum_{l=1}^r\frac{\car_{kl}}{2}\int_{-\mu_l}^{\mu_l}dy \rho_l(y)\left(\theta(x-y)-\theta(x)\right)=\frac{\sqrt{\mu_k^2-x^2}}{2\pi g_k^2}\nonumber\\&-\frac{1}{\pi}\sum_{l=1}^r\frac{\car_{kl}}{2}\int_{|w|>\mu_k}\frac{\text{sgn}(w)dw}{w-x}\sqrt{\frac{\mu_k^2-x^2}{w^2-\mu_k^2}}\int_{-\mu_l}^{\mu_l}dy\rho_l(y)(\theta(w-y)-\theta(w))=:\mathbb{F}_k(x)\,.
\end{align}
The functions $\mathbb{F}_k$ are the \textit{driving terms} of the integral equations, i.e. our equations are written as
\beq
\label{eq:cyclicquiverintegralequations}
\rho_k(x)-\sum_{l=1}^\nv\frac{\car_{kl}}{2}\int_{-\mu_l}^{\mu_l}dy \rho_l(y)\left(\theta(x-y)-\theta(x)\right)=\mathbb{F}_k(x)\,.
\eeq
In the strong coupling limit, the $\mathbb{F}_k(x)$ are dominated by their first term.
Fourier transforming \eqref{eq:cyclicquiverintegralequations}, using \eqref{eq:fourriertranformoftheta} and $\sum_{l=1}^\nv\car_{kl}=0$, we get the set of equations
\beq
\label{eq:strongcouplingFourier}
\hat{\rho}_k(\omega)+\sum_{l=1}^\nv\frac{\car_{kl}}{2}\frac{\hat{\rho}_l(\omega)}{2\sinh^2\frac{\omega}{2}}=\hat{\mathbb{F}}_k(\omega)+e^{-i\mu_k \omega}\hat{X}_k^-(\omega)+e^{i\mu_k \omega}\hat{X}_k^+(\omega),
\eeq
where the functions $\hat{X}_k^{\pm}$ are analytic in the upper/lower half plane respectively. Their presence is due to the fact that \eqref{eq:cyclicquiverintegralequations} only holds for $x\in(-\mu_k,\mu_k)$. 
We can now rewrite \eqref{eq:strongcouplingFourier} as
\beq
\label{eq:strongcouplingFourier2}
\sum_{l=1}^\nv \mathbb{A}_{kl}(\omega)\hat{\rho}_l(\omega)=\hat{\mathbb{F}}_k(\omega)+e^{-i\mu_k \omega}\hat{X}_k^-(\omega)+e^{i\mu_k \omega}\hat{X}_k^+(\omega)
\eeq
where 
\beq
\mathbb{A}_{kl}(\omega)\colonequals \frac{(e^{\omega}+e^{-\omega})\delta_{kl}-\delta_{k,l-1}-\delta_{k,l+1}}{e^{\omega}+e^{-\omega}-2}\,
\eeq
with the indices $k,l$ subject to the identification $k\equiv k+\nv$ and $l\equiv l+\nv$. The implicit dependence on the number of gauge groups $\nv$ contained in many quantities, such as $\mathbb{A}_{kl}$, will not be explicitly noted in this appendix so as to not clutter the notation.
Solving \eqref{eq:strongcouplingFourier2} for the densities by inverting the matrix $\mathbb{A}$, we get
\beq
\begin{split}
\label{eq:strongcouplingFourierinverted}
\mathbb{K}(\omega)\hat{\rho}_k(\omega)=&\sum_{l=1}^\nv\left(\chi_{|k-l|-1}(e^\omega)+\chi_{\nv-1-|k-l|}(e^\omega)\right)\\&\times \left(\hat{\mathbb{F}}_l(\omega)+e^{-i\mu_l \omega}\hat{X}_l^-(\omega)+e^{i\mu_l \omega}\hat{X}_l^+(\omega)\right)\,,
\end{split}
\eeq
where $\chi_j(u)\colonequals \sum_{m=0}^ju^{j-2m}$ are the characters of the $j+1$ dimensional representations of $\SU{2}$ and we have defined the kernel
\beq
\mathbb{K}(\omega)\colonequals  \frac{e^{\nv\omega}-2+e^{-\nv\omega}}{e^{\omega}-2+e^{-\omega}}=\left(\frac{\sinh\frac{\nv\omega}{2}}{\sinh\frac{\omega}{2}}\right)^2\,.
 \eeq 
 Using the well known formula $\sin(x)=\pi\left(\Gamma(\frac{x}{\pi})\Gamma(1-\frac{x}{\pi})\right)^{-1}$, we can express $\mathbb{K}(\omega)$ using the $\Gamma$ function as
 \beq
 \mathbb{K}(\omega)=\frac{1}{\mathbb{G}_+(\omega)\mathbb{G}_-(\omega)}, \quad \text{ where } \quad \mathbb{G}_{\pm}(\omega)\colonequals \frac{1}{\nv}\left(\frac{\Gamma\big(1\mp \frac{i\nv\omega}{2\pi}\big)}{\Gamma\big(1\mp \frac{i\omega}{2\pi}\big)}\right)^2\,,
 \eeq
 where the functions $ \mathbb{G}_{\pm}(\omega)$ have second order poles at $\mp i\nu_n$ with 
 \beq
 \nu_n=2\pi \frac{n}{\nv}, \qquad n=1,2,3,\ldots, \text{ with } n\notin \nv\mathbb{N}.
 \eeq
Expanding around the poles, we get 
\beq
\mathbb{G}_{\pm}(\omega)= \frac{\alpha_n}{(\omega \pm i\nu_n)^{2}}\pm \frac{\beta_n}{\omega \pm i\nu_n}+\cdots
\eeq 
where 
 \beq
 \label{eq:coefficientsogGplus}
 \alpha_n\colonequals -\frac{4 \pi^2}{\nv^3\Gamma(n)^2\Gamma\big(1-\frac{n}{\nv}\big)^2}\,, \qquad \beta_n\colonequals\frac{4\pi i\left(r\psi(n)-\psi\big(1-\frac{n}{\nv}\big)\right)}{\nv^3\Gamma(n)^2\Gamma\big(1-\frac{n}{\nv}\big)^2}\,,
 \eeq
 and $\psi(x)\colonequals \nicefrac{\Gamma'(x)}{\Gamma(x)}$ is the digamma function. Observe that both $\alpha_n$ and $\beta_n$ go to zero very rapidly and that the $\beta_n$ are purely imaginary. 
We solve \eqref{eq:strongcouplingFourierinverted} by multiplying by $\mathbb{G}_+(\omega)e^{-i\mu_k\omega}$ and taking the negative frequency part. Ignoring to first approximation the $\hat{X}^{\pm}_k$, we get
\beq
\label{eq:densitiesnegativefrequencies}
\hat{\rho}_k(\omega)=\mathbb{G}_-(\omega)e^{i\mu_k\omega}\left[\mathbb{G}_+(\omega)e^{-i\mu_k\omega}\sum_{l=1}^\nv\mathbb{B}_{kl}(\omega)\hat{\mathbb{F}}_l(\omega)\right]_-\,,
\eeq 
where we used the matrix $\mathbb{B}$ with
\beq
\mathbb{B}_{kl}(\omega)\colonequals \chi_{|k-l|-1}(e^{\omega})+\chi_{\nv-1-|k-l|}(e^{\omega})\,
\eeq
and the definition
\beq
\label{eq:projectionfrequencies}
\mathcal{F}_{\pm}(\omega)\colonequals \pm \lim_{\epsilon\rightarrow 0+}\int_{-\infty}^{\infty}\frac{d\omega'}{2\pi i }\frac{\mathcal{F}(\omega')}{\omega'-\omega\mp i\epsilon}\,.
\eeq
The functions $\mathcal{F}_{\pm}(\omega)$ are analytic in the upper/lower half-planes and the integral contours are to be closed in the upper/lower half-plane. The contour integral of \eqref{eq:densitiesnegativefrequencies} will give us the residues of $\mathbb{G}_+(\omega)\mathbb{B}_{kl}(\omega)$ at $-i\nu_n$.
Expanding, we find 
\beq
\text{Res}\left(\mathbb{G}_+(\omega)\mathbb{B}_{kl}(\omega), -i\nu_n\right)=\alpha_n(\partial_{\omega}\mathbb{B}_{kl})(-i\nu_n)+\beta_n\mathbb{B}_{kl}(-i\nu_n)\,.
\eeq
Through brute force, we discover that
\beq
\label{eq:coefficientschi}
\mathbb{B}_{kl}(-i\nu_n)=\left\{\begin{array}{ll}r &\text{ for }  n\in r\mathbb{N}\\(-1)^{k+l-1}r & \text{ for } n\in \frac{r}{2}\mathbb{N}\\0&\text{ otherwise}\end{array}\right.\,.
\eeq
as well as
\beq
\label{eq:derivativechi}
(\partial_\omega \mathbb{B}_{kl})(-i\nu_n)=\left\{\begin{array}{ll}0 &\text{ for }  n\in r\mathbb{N}\\0& \text{ for } n\in \frac{r}{2}\mathbb{N}\\\frac{ir \cos(|k-l|\nu_n)}{\sin(\nu_n)}&\text{ otherwise}\end{array}\right. \,.
\eeq
Using the above, we find the residue of the poles $-i\nu_n$:
\beq
\label{eq:ResGpB}
\mathbb{R}_{kl}(n)\colonequals \text{Res}\left(\mathbb{G}_+(\omega)\mathbb{B}_{kl}(\omega), -i\nu_n\right)=\left\{\begin{array}{ll} 0 & \text{ for } n\in r\mathbb{N},\\\beta_n(-1)^{k+l-1}r & \text{ for } n\in \frac{r}{2}\mathbb{N},\\ \alpha_n \frac{ir \cos(|k-l|\nu_n)}{\sin(\nu_n)}& \text{ otherwise}\end{array}\right.
\eeq
We observe experimentally that the matrices $\mathbb{R}_{kl}(n)$ all commute, that the vector $(1,1,\ldots,1)^t$ is the only common eigenvector of $\mathbb{R}_{kl}(n)$ and that it has eigenvalue zero.
Plugging \eqref{eq:projectionfrequencies} into  \eqref{eq:densitiesnegativefrequencies}  we get
\beq
\label{eq:densitiesnegativefrequencies2}
\hat{\rho}_k(\omega)=\frac{1}{\mathbb{K}(\omega)}\sum_{l=1}^r\mathbb{B}_{kl}(\omega)\hat{\mathbb{F}}_l(\omega)-\mathbb{G}_-(\omega)e^{i\mu_k\omega}\sum_{n=1}^{\infty}\frac{e^{-\mu_k\nu_n}}{\omega+i\nu_n}\sum_{l=1}^r\mathbb{R}_{kl}(n)\hat{\mathbb{F}}_l(-i\nu_n)\,,
\eeq
where we have to use  \eqref{eq:ResGpB} for the residues. Observe that, contrary to what one might think at first glance, \eqref{eq:densitiesnegativefrequencies2} is indeed analytic in the lower half plane since the poles at $-i\nu_n$ cancel.

We now need to normalize the densities. Since the Fourier transform of \eqref{eq:densitiesnegativefrequencies2} will represent the density $\rho_k(x)$ well only for positive $x$, we use the fact that the densities should be symmetric and demand
\beq
1=2\int_0^{\mu_k}dx\rho_k(x)=\lim_{\epsilon\rightarrow 0+}\int_{-\infty}^{\infty}\frac{d\omega}{\pi i}\frac{\hat{\rho}_k(\omega)}{\omega-i\epsilon}=2(\hat{\rho}_k)_+(0)\,,
\eeq
where we used \eqref{eq:projectionfrequencies}. Thus, we need to close the contour in the upper half plane.

Taking the residues and using $\frac{1}{\omega\mp i \epsilon}=\text{p.v.}\pm i \pi \delta$ as well as the symmetry under $\omega\rightarrow -\omega$ of $\mathbb{B}$, $\mathbb{K}$ and $\hat{\mathbb{F}}_k$ gives
\beq
\begin{split}
1=&\frac{1}{\mathbb{K}(0)}\sum_{l=1}^r\mathbb{B}_{kl}(0)\hat{\mathbb{F}}_l(0)+2i\mathbb{G}_-(0)\sum_{n=1}^{\infty}\frac{e^{-\mu_k\nu_n}}{\nu_n}\sum_{l=1}^r\mathbb{R}_{kl}(n)\hat{\mathbb{F}}_{l}(-i\nu_n)\\
&-2\sum_{m,n=1}^{\infty}\frac{e^{-\mu_k(\nu_m+\nu_n)}}{\nu_m(\nu_m+\nu_n)}\left(\beta_m-i\alpha_m\frac{\nu_n+2\nu_m+\nu_n\nu_m\mu_k+\nu_m^2\mu_k}{\nu_m(\nu_m+\nu_n)}\right)\sum_{l=1}^r\mathbb{R}_{kl}(n)\hat{\mathbb{F}}_{l}(-i\nu_n)
\end{split}
\eeq
Using $\mathbb{K}(0)=r^2$, $\mathbb{B}_{kl}(0)=r$ and $\mathbb{G}_-(0)=\frac{1}{r}$, we get
\beq
\label{eq:finalstrongcouplingequations}
\begin{split}
1=&\frac{1}{r}\sum_{l=1}^r\hat{\mathbb{F}}_l(0)+\frac{2i}{r}\sum_{n=1}^{\infty}\frac{e^{-\mu_k\nu_n}}{\nu_n}\sum_{l=1}^r\mathbb{R}_{kl}(n)\hat{\mathbb{F}}_{l}(-i\nu_n)\\
&-2\sum_{m,n=1}^{\infty}\frac{e^{-\mu_k(\nu_m+\nu_n)}}{\nu_m(\nu_m+\nu_n)}\left(\beta_m-i\alpha_m\frac{\nu_n+2\nu_m+\nu_n\nu_m\mu_k+\nu_m^2\mu_k}{\nu_m(\nu_m+\nu_n)}\right)\sum_{l=1}^r\mathbb{R}_{kl}(n)\hat{\mathbb{F}}_{l}(-i\nu_n)\,.
\end{split}
\eeq
The terms in the second line of \eqref{eq:densitiesnegativefrequencies2} are exponentially suppressed. We will now make an assumption that is justified by the self-consistency of the results. As in \cite{Passerini:2011fe}, in the large coupling limit, the driving terms defined in \eqref{eq:definitionofthedrivingterm} are dominated by their first term, so that
 \beq
 \mathbb{F}_k(x)=\frac{1}{2\pi g_k^2}\sqrt{\mu_k^2-x^2}\Longrightarrow \hat{\mathbb{F}}_k(\omega)=\frac{\mu_kJ_1(\mu_k\omega)}{2g^2_k\omega},\quad \hat{\mathbb{F}}_k(0)=\frac{\mu_k^2}{4g^2_k}\,,\quad \hat{\mathbb{F}}_k(-i\nu_n)=\frac{e^{\mu_k\nu_n}}{g_k^2}\sqrt{\frac{\mu_k}{8\pi\nu_n^3}}\,,
  \eeq 
where $J_n$ are the Bessel functions of the first kind. Using this approximation for the driving terms and dropping the exponentially suppressed terms in \eqref{eq:finalstrongcouplingequations},  explicit numerical solutions show that for large values of the couplings, the widths of the densities $\mu_k$ are very close to being equal to each other (as long as the ratios of the couplings $\nicefrac{g_k}{g_l}$ is roughly of order one) and are to a very good approximation given by 
\beq
\label{eq:strongcouplinglimitofmu}
  \mu_k=\bar{\mu}\colonequals 2\sqrt{r\left(\sum_{k=1}^rg_{k}^{-2}\right)^{-1}}=2\sqrt{\frac{rg_1^2\cdots g_r^2}{\sum_{i=1}^r\prod_{k\neq i}g_k^2}}\qquad \forall k\,.
  \eeq
It can be seen that this is a solution of \eqref{eq:finalstrongcouplingequations} if we ignore the (numerically suppressed) pieces containing $\hat{\mathbb{F}}_k(-i\nu_n)$, which implies
\beq
1=\frac{1}{\mathbb{K}(0)}\sum_{l=1}^r\mathbb{B}_{kl}(0)\hat{\mathbb{F}}_l(0)=\frac{1}{r^2}\sum_{l=1}^rr\frac{\mu_l^2}{4g^2_l}=\frac{1}{4r}\sum_{l=1}^r\frac{\mu_l^2}{g_l^2}\,.
\eeq 
An a posteriori justification for neglecting the $\hat{\mathbb{F}}_k(-i\nu_n)$ is that, for $g_k$  that are not wildly different, see \eqref{eq:scalingofthecouplings}, the $\hat{\mathbb{F}}_k$ will be roughly equal, so that the term $\sum_{l=1}^r\mathbb{R}_{kl}(n)\hat{\mathbb{F}}_{l}(-i\nu_n)$ will be small since $(1,1,\ldots, 1)$ is an eigenvector of the $\mathbb{R}$ matrices with eigenvalue zero.

The densities themselves have in the strong coupling limit the same shape as the $\calN=4$ one. The leading behavior of the Wilson loop expectation  values at large values of the coupling is hence (using the asymptotic expression $I_1(x)\sim \nicefrac{e^{x}}{\sqrt{2\pi x}}$)
\beq
\langle W_k^{\pm}\rangle=\hat{\rho}_k(2\pi i b^{\pm 1})\sim \frac{ e^{2 \pi  b^{\pm 1} \bar{\mu} }}{2 \pi ^2 b^{\pm 3/2} \bar{\mu}^{\frac{3}{2}}}+\mathcal{O}(b-1)^2\,.
\eeq
Due to the exponential term, the strong coupling limit of the effective couplings is simply given by comparing \eqref{eq:strongcouplinglimitofmu} with the width $\mu$ for $\calN=4$. Since for $\calN=4$ SYM we have $\mu^2=4g^2$, comparing with \eqref{eq:strongcouplinglimitofmu} leads to equation \eqref{eq:fkinthestrongcoupling} in the main text.  In particular, the strong coupling limit of the Bremsstrahlung functions is 
\beq
B_k=\frac{\mu_k }{2 \pi }-\frac{3}{8 \pi ^2}\,.
\eeq
A more detailed analysis of \eqref{eq:finalstrongcouplingequations} should also allow one to extract the leading corrections to the relation \eqref{eq:fkinthestrongcoupling}.


\providecommand{\href}[2]{#2}\begingroup\raggedright\endgroup

\end{document}